\documentclass[twocolumn]{aastex6}
\usepackage{amsmath}
\newcommand{\mdot}{\dot{\mathcal{M}}}
\newcommand{\psimin}{\Psi_{\rm min}}
\newcommand{\tratio}{\tau_{\rm adv}/\tau_{\rm heat}}
\newcommand{\rnu}{R_{\nu}}

\newcommand{\epssclone}{\epsscale{1.0}}
\newcommand{\epsscltwo}{\epsscale{1.0}}
\newcommand{\epssclthree}{\epsscale{1.0}}
\newcommand{\epssclfour}{\epsscale{1.0}}

\shorttitle{}
\shortauthors{}

\begin{document}

\title{An Integral Condition for Core-Collapse Supernova Explosions}

\author{Jeremiah W. Murphy\altaffilmark{1}}
\author{Joshua C. Dolence\altaffilmark{2}}

\altaffiltext{1}{Florida State University, Tallahassee, FL, U.S., jwmurphy@fsu.edu}
\altaffiltext{2}{CCS-2, Los Alamos National Laboratory, P.O. Box 1663,
Los Alamos, NM 87545}

\begin{abstract}
We derive an integral condition for core-collapse supernova (CCSN)
explosions and use it to construct a new diagnostic of explodability.  
The fundamental challenge in CCSN theory is to
explain how a stalled accretion shock revives to explode a star.
In this manuscript, we assume that the shock revival
is initiated by the delayed-neutrino mechanism and derive an integral
condition for spherically symmetric shock expansion, $v_s > 0$.
One of the most useful one-dimensional explosion conditions is the
neutrino luminosity and mass-accretion rate ($L_{\nu}$-$\mdot$) critical
curve.  Below this curve, steady-state stalled
solutions exist, but above this curve, there are no
stalled solutions.  \citet{burrows93} suggested that the
solutions above this curve are dynamic and explosive.  In this manuscript, we 
take one step closer to proving this supposition; we show that {\it
all} steady solutions above this curve have $v_s > 0$.  Assuming that
these steady $v_s > 0$ solutions correspond to explosion, we present a
new dimensionless integral condition for explosion, $\Psi > 0$.
$\Psi$ roughly describes the balance between
pressure and gravity, and we show that this parameter is equivalent to
the $\tau$ condition used to infer the $L_{\nu}$-$\mdot$ critical
curve. The illuminating difference is that there is a direct
relationship between $\Psi$ and $v_s$.  Below the critical curve,
$\Psi$ may be negative, positive, and zero, which corresponds to
receding, expanding, and stalled-shock solutions.  At the critical
curve, the minimum $\Psi$ solution is zero; above the critical curve,
$\psimin > 0$, and all steady solutions have $v_s > 0$.  Using one-dimensional
simulations, we confirm our primary assumptions and verify that
$\psimin > 0$ is a reliable and accurate explosion diagnostic.

\end{abstract}


\section{Introduction}
\label{sec:introduction}

The question of how typical massive stars explode as core-collapse
supernovae (CCSNe) has plagued theorists for decades \citep{colgate66}.  The collapse and bounce of the Fe core of
a massive star launches a strong shock wave; however, this prompt
shock quickly stalls as a result of electron capture,
nuclear dissociation, and neutrino emission \citep{hillebrandt81,mck82,mazurek82}.  Thus the explosive shock
is momentarily aborted, producing a stalled accretion shock.  If the shock remains stalled, the explosion will fail and the proto-neutron star will
continue to accrete, eventually collapsing to a black hole
\citep{fischer09,oconnor11}.  We know, however, that stars
explode and many times leave neutron stars \citep{li2010,horiuchi2011,fryer12}.
Therefore, the fundamental question in core-collapse theory is how
a CCSN transitions from a stalled accretion shock phase into a
phase of runaway shock expansion that explodes the star.  In this
paper, we derive an integral condition in the limiting case
  of spherical symmetry that divides
stalled-shock solutions from explosive solutions.

The prevailing view is that neutrinos help to reinvigorate the
stalled shock, leading to runaway expansion \citep{bethe85,janka12,burrows13}.
A large neutrino flux cools the proto-neutron star and surrounding
regions; about half of that luminosity comes from neutrinos that diffuse
out of the natal proto-neutron star, and the rest is emitted directly as
accretion luminosity by
the accreting stellar material.  A fraction of these neutrinos recapture just
below the shock, depositing enough energy in the post-shock mantle to
reinvigorate the stalled accretion shock and initiate explosion.
However, when this picture is simulated in numerical models with
significant (although perhaps not sufficient) detail, the final and most
important element --- explosion --- remains elusive or, at best, inconsistent 
with observations \citep{ott08,muller12,hanke13,takiwaki14,lentz15,dolence15,melson2015,bruenn2016,roberts2016}.  Therefore, a major
challenge for simulators has been to understand the various physical
effects that influence the unsatisfactory outcomes, all in the context
of enormously complicated and expensive simulations.  This process of
interpretation has inevitably relied on some easily understood models
that, one hopes, encapsulate the important physics of the problem.

In this context, perhaps the most impactful model was proposed by
\citet{burrows93}.  They proposed that the essence of the core-collapse
problem is captured by considering a simple boundary value problem.
The inner boundary is set at the proto-neutron star ``surface,''
defined by a neutrino optical depth of 2/3, where a temperature and
therefore luminosity is specified.  The outer boundary is at the
stalled shock, where the post-shock solutions match the nearly free-falling stellar material upstream via the jump conditions.
\citet{burrows93} proceeded to show that hydrodynamic solutions with a
stationary shock only exist below a critical luminosity and accretion-rate curve.  Above this curve, no stalled-shock solutions exist.  They
speculated that this critical curve, separating stationary from
non-stationary solutions, also represents a critical curve for
explosion; they suggested (but did not prove) that
  non-stationary solutions are explosive.  Subsequent work with parameterized one-, two-, and three-dimensional core-collapse models have
confirmed qualitatively that such a critical curve exists
\citep{murphy08b,hanke12,dolence13,couch13}, but its
precise location in parameter space depends on other details of the
problem, including, e.g., the structure of the progenitor, which hampers
the use of the critical curve in practice \citep{suwa14,dolence15}.

Nonetheless, the idea of criticality has framed much of the discussion
surrounding the simulation results and has motivated the introduction
of heuristic and approximate measures of ``nearness to explosion.''
For example, many have suggested that the ratio of the advection
timescale\footnote{Time to advect through the net heating region} to
heating timescale\footnote{Time to significantly change the thermal
  energy in the net heating region} captures an important aspect of the
problem, with values $\gtrsim$1 conducive to explosion 
\citep{janka98,thompson00,thompson05,buras06b,murphy08b}.  Similarly, \citet{pejcha12} argue that their
``antesonic'' condition represents a critical condition for explosion.
Both conditions, however, suffer the same afflictions: they lack
precise critical values, and they both run away only after explosion
commences.  These shortcomings have led to the practice of measuring
these parameters as a function of simulation time and deciding on
``critical values'' ex post facto \citep{muller12,dolence13,couch13}.  Clearly, this is
unsatisfactory and leads to little more insight than simply
identifying explosions with runaway shock radii.  In principle, these
critical values may even be reached precisely \textit{because} of
explosion; the relationship may be symptomatic rather than causal.

Investigating criticality has proven to be a
useful quantitative measure of explodability.  For some time, it was
very clear that the delayed-neutrino mechanism fails in one-dimensional,
spherically symmetric simulations \citep{liebendorfer01a,liebendorfer01b,liebendorfer05b,rampp02,buras03,thompson03}; it was also becoming apparent that
multi-dimensional simulations showed promise where one-dimensional simulations failed \citep{herant94,janka95,janka96,burrows95,burrows07d,melson2015,bruenn2016,roberts2016}.  The
critical curve offers a quantitative measure of how much the
multidimensional instabilities aide the neutrino mechanism toward
explosion.  \citet{murphy08b} investigated in one and two dimensions whether a
critical curve is even sensible in time-dependent simulations.  They not only
empirically found that a critical curve is viable, but that it is
about 30\% lower in two-dimensional simulations.  Subsequent investigations indicate
that the critical curve in three-dimensional is similar to two-dimensional simulations \citep{couch12,hanke13}.  There are
certainly slight differences between two and three dimensions, but these differences are minor
in comparison to the major shift that multidimensional instabilities
enable in going from one dimension to multidimensionality.  Further investigations suggest
that turbulence plays a major role in reducing the critical condition
for explosion \citep{murphy11b,murphy13}.  While multidimensionality and turbulence are
important considerations in the explosion mechanism, we do not
address turbulence in this paper; we leave that for a subsequent paper
(Q.~Mabanta et al., in preparation).  In this manuscript, we aim to further understand the
foundational aspects of criticality, and later, we hope to expand
these to include turbulence.

In this work, we revisit the idea of criticality, generalizing the
critical-curve concept to a critical hypersurface that depends on
all of the relevant parameters.  We show that for a fixed set of
parameters, there is a
family of possible solutions.  Depending upon the parameters, each family falls
into one of two categories.  In one category, the family consists of
solutions with negative, zero, or positive shock
velocity.  For such a family, the zero-shock-velocity
solution is the quasi-stationary solution.  In the second category, all
possible solutions have positive shock velocity.  These two
categories are divided by a critical hypersurface, which may be expressed as a single dimensionless parameter.

Our initial motivation in revisiting criticality was to ask why does
the neutrino-luminosity and accretion-rate curve of \citet{burrows93}
corresponds to explosion in simulations.  Under certain restrictive but
empirically reasonable assumptions, we derive that the only possible
solutions above the curve correspond to positive shock velocity.  Not
only does this derivation suggest a reason for explosion, but it also
suggests a more general critical condition.

This new condition for explosion, the critical hypersurface and
associated dimensionless parameter, proves to be a useful diagnostic for
core-collapse simulations.  For one, we empirically show that the
hypersurface corresponds to the transition to explosion in
parameterized one-dimensional CCSN models.
Further, the associated dimensionless parameter reliably and quantitatively indicates
when explosions commence.  Since we derive the parameter directly from the
equations of hydrodynamics,  its usefulness is not limited by the ad hoc calibrations of other popular measures.  Finally, when combined with
semi-analytic models, we show that our single parameter yields an
accurate and reliable measure of nearness-to-explosion.

For your convenience, the structure of this manuscript is as follows.
In section~\ref{sec:boundaryproblem}, we review the boundary value
problem that describes the core-collapse problem \citep{burrows93} and identify the important parameters
of the problem.  With this framework, and with the proposition that $v_s >
0$ corresponds to explosion, we derive the integral condition for
explosion in section~\ref{sec:integralcondition}.  Then in section~\ref{sec:validate}, we validate the
integral conditions for explosion with one-dimensional parameterized simulations.
In section~\ref{sec:explodability}, we use the integral condition to investigate the
family of steady-state solutions and propose an explosion
diagnostic.  Then in section~\ref{sec:comparison}, we compare the reliability of the integral
condition explosion diagnostic with other popular explosion measures, finding
that the integral condition outperforms them all.  Finally, for a
brief summary of the conclusions, cautionary notes, and future
prospects, see section~\ref{sec:conclusions}.


\section{Quasi-steady Solutions: A Boundary Value Problem}
\label{sec:boundaryproblem}

The fundamental question of core-collapse theory is how does the stalled accretion shock phase transitions into a dynamic explosion.  In other words, what are the conditions for which the
shock velocity, $v_s$, is persistently greater than zero?
While the shock is stationary, the steady-state assumption is quite
good.  Under this assumption, the entire region below the shock may be
treated as a boundary value problem \citep{burrows93}.  The upper boundary is set by the properties of the nearly free-falling stellar material and the jump conditions at the shock, while the lower boundary is
the surface of the neutron star.  Later, we derive a condition in which
there are no more $v_s = 0$ solutions, and the only solutions left are
those in which $v_s > 0$.  To understand when the steady-state
solutions are no longer viable, we must first understand the stalled
accretion shock solution and the important parameters of the problem.

To begin, the governing conservation equations are
\begin{equation}
\label{eq:mass}
\frac{\partial \rho}{\partial t} + \nabla \cdot (\rho {\bf v}) = 0 \, ,
\end{equation}
\begin{equation}
\label{eq:momentum}
\frac{\partial (\rho {\bf v})}{\partial t} + \nabla \cdot (\rho {\bf v}
     {\bf v}) + \nabla P = \rho \nabla \phi \, ,
\end{equation}
and
\begin{equation}
\label{eq:energy}
\frac{\partial (\rho E) }{\partial t} 
+ \nabla \cdot \left [ \rho {\bf v} \left ( h +
  \frac{v^2}{2} \right )  \right ] = \rho {\bf v}
\cdot \nabla \phi + \rho q \, ,
\end{equation}
where $\rho$ is the mass density, ${\bf v}$ is the velocity, $P$ is
the pressure, $\phi$ is the gravitational potential, 
$E = \varepsilon + v^2/2$ is the internal plus kinetic specific energy, and
$h = \varepsilon + P/\rho$ is the specific enthalpy.  In this paper,
we approximate the heating and cooling, $q$ as
\begin{equation}
\label{eq:heatcool}
q = \frac{L_{\nu}\kappa}{4 \pi r^2} - C \left ( \frac{T}{T_0} \right
)^6 \, .
\end{equation}
The first term is a model in which we treat the neutrino heating as if
all neutrinos were emitted from the proto-neutron star ``surface'' with
a luminosity of $L_{\nu}$ \citep{janka01,murphy08b,murphy13}.  $\kappa$ is the opacity for absorbing
neutrinos in the region above the ``surface,''
which is proportional to $T_{\nu}^2$.  In this paper, the surface of
the proto-neutron star is given by $\tau = \int_{\rnu}^{\infty} \kappa \rho \, dr =
2/3$, which implicitly defines the proto-neutron star radius ($\rnu$) as the neutrinospheric radius.  In practice, we find that this corresponds to a density of
$\rho_{\rm 2/3} \approx 7 \times 10^{10}$ g cc$^{-1}$.  The second term is the neutrino
cooling due to thermal weak interactions \citep{janka01}.

The neutrino interactions that we employ are quite simple, and
  while they are crude, they retain some key elements
  allowing us to more easily assess the importance and role of
  neutrino heating and cooling.  In the simple model, there are three
  parameters to describe the neutrino heating, the core neutrino
  luminosity, the temperature of the neutrinos, and the
  neutrino-sphere radius.  A better description of the neutrino luminosity would include an accretion luminosity, which would account for the
  added neutrino luminosity provided by the semi-transparent cooling
  region above the neutrino sphere.  While it is not difficult to
  include such an accretion luminosity \citep{pejcha12}, we decided to
  employ the more traditional ``light bulb'' description to reduce
  the confounding prescriptions in our initial investigation.  Later,
  we will include the accretion luminosity and investigate the
  differences and whether they are merely quantitative or are more
  substantially qualitative.  Even though the neutrino spectrum is
  not quite Planckian, we treat it as such to reduce the neutrino energy
  spectrum to one parameter, the neutrino temperature.  The final
  parameter, the neutrino sphere radius, is probably even more
  approximate.  Even if one would be able to describe the neutrino
  spectrum with one temperature, there is a distribution of neutrino
  energies, and since the cross section depends upon the square of the
  neutrino energy, the neutrino sphere for each energy would occur at
  a different radius.  Nonetheless, we seek a parameterization of the
  size of the proto-neutron star and choosing one neutrino sphere
  radius which emits the core luminosity at a core temperature probably
  preserves the qualitative behavior.   Given these
  differences, we present the ideas and results of this paper as
  qualitative arguments.  We suspect that subsequent more realistic models for
  neutrino interactions will preserve the qualitative nature of our
  conclusions.

In a quasi-steady-state, the time-derivative terms are small and the
problem is well-approximated by setting these terms to zero.
We specifically refer to the quasi-steady-state because we are
  assuming that the post-shock profile is steady, but the shock
  velocity is nonzero.  Often when one assumes steady-state, one also assumes that the
  shock velocity is zero.  However, it is possible to have a quasi-steady nonlinear solution
  behind the shock and a nonzero shock velocity.  The Noh test problem
  is an example of one such nonlinear solution.  In this test problem,
  a supersonic flow is incident on a wall.  A shock forms, and the
  post-shock flow maintains the same density profile; i.e. it is
  steady.  However, the shock has a nonzero lab frame velocity.
  Similarly, in finding quasi-steady solutions, we assume that the
  post-shock flow is steady (i.e. the time-derivative terms are zero),
  but we explicitly allow for a {\it nonzero} shock velocity.

\begin{figure*}[t]
\plotone{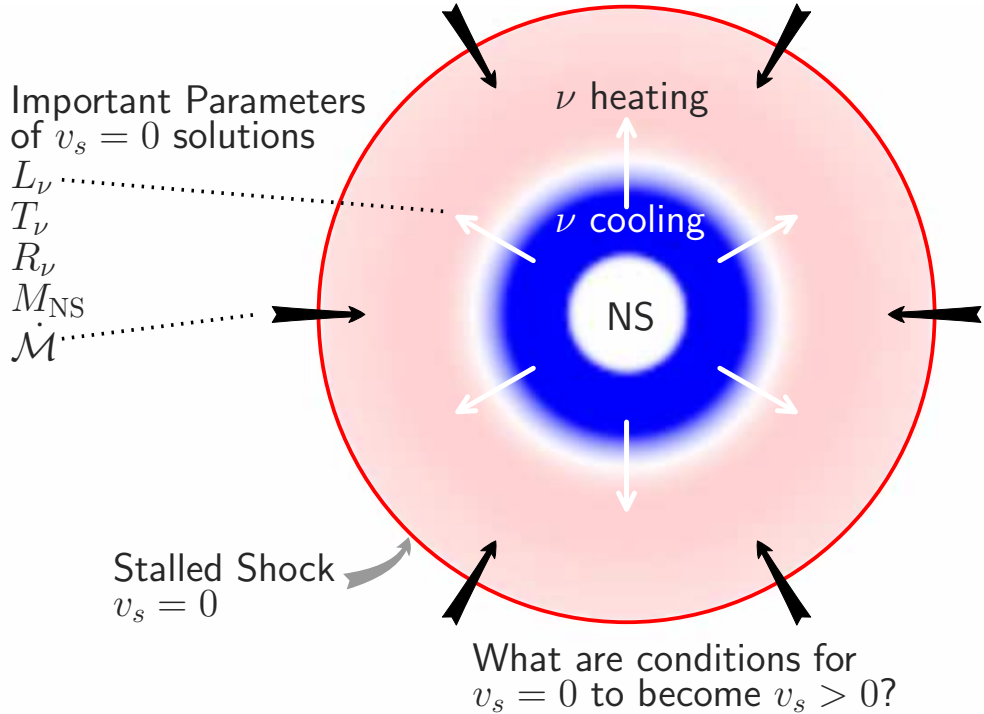}
\caption{After the bounce shock stalls, the post-shock flow settles into a quasi-steady configuration.  The solutions satisfy a boundary value problem between the neutron star ``surface'' and the stalled shock.  The
  important parameters in determining the steady-state solutions are
  the neutrino luminosity being emitted from the core ($L_{\nu}$), the
  temperature of the neutrinos ($T_{\nu}$), the mass-accretion rate
  onto the shock and NS ($\mdot$), the neutrino sphere or NS radius ($\rnu$), and
  the NS mass ($M_{\rm NS}$).  The fundamental challenge in
  core-collapse theory is to understand how $v_s = 0$ transitions to
  $v_s > 0$.  In section~\ref{sec:integralcondition}, we derive such a condition. \label{fig:boundaryvalue}}
\epssclone
\end{figure*}

The solution to the resulting quasi-steady boundary value problem describes the
conditions of the flow between the proto-neutron star surface and the
bounding shock in terms of the important parameters of the
problem: $L_{\nu}$, $T_{\nu}$, $\rnu$, $M_{\rm NS}$
(proto-neutron star mass), and $\dot{M}$ (accretion rate).
Even though we have assumed that the post-shock flow is steady,
the shock velocity may or may not be zero.  We ask the fundamental
question what it takes for all quasi-steady-state solutions to have
$v_s > 0$.  Later, we will show that
for certain values of these important parameters {\it all} of the
quasi-steady state solutions have $v_s > 0$.  See
Figure~\ref{fig:boundaryvalue} for a schematic of the boundary value
problem and the most important parameters of the problem.

The five parameters that we highlight are a natural
  parameterization for the core-collapse problem and offer a way to
  parameterize some of the most uncertain aspects of the core-collapse
  problem.  For example, two uncertain aspects of the core-collapse
  problem are the structure of the progenitor and the dense nuclear
  equation of state (EOS).

The uncertainties of the dense nuclear EOS for neutron stars are often parameterized in terms of a
  mass-radius relationship \citep{lattimer2016}.  Each EOS predicts a specific
  mass-radius curve.  Therefore, investigating how
  the conditions for explosion depend upon the proto-neutron
  star mass, $M_{\rm NS}$, and the neutrino-sphere radius (a proxy for
  the neutron star radius), $\rnu$, provides a means to parameterize how
  the condition for explosion depends upon the uncertainties in the
  EOS.

The three other parameters help to parameterize the
  uncertainties in the progenitor structure as well as neutrino
  diffusion in the core.  For example, the time evolution of both
  $M_{\rm NS}$ and $\dot{M}$ depends upon the progenitor structure.
  $L_{\nu}$ depends upon a mix of the total thermal energy available
  in the neutron structure, neutrino diffusion, and the accretion
  rate.  Hence, $L_{\nu}$ is not entirely independent of the other
  parameters.  In a subsequent paper, we will explore the consequences
  of these extra constraints.  For now, however, we use $L_{\nu}$ to discuss our
  new explosion condition in the context of previous conditions.  The neutrino
  temperature is also not entirely independent (e.g. 
$L_{\nu} \approx 4 \pi R_{\nu}^2 \sigma T_{\nu}^4$), but again we use it to make a connection
  to past literature.  In a future paper, in which we discuss
  analytic solutions, we will propose an alternative formulation for
  these parameters.  Until then, these particular parameters are a natural parameterization of the
core-collapse problem that also incorporate the uncertainties in the
progenitor and neutron star physics.

\section{Deriving an Integral Condition for $v_s > 0$}
\label{sec:integralcondition}

We reduce the core-collapse problem to a set of integral conditions
and show that this leads to a critical condition for explosions.
Before we derive the integral conditions, let us describe what
motivated us to consider
the integral equations at all.  The governing equations
(eqs.~\ref{eq:mass}-\ref{eq:energy}) are commonly presented in differential form, but
one could just as easily present them in integral form.  The two forms
represent exactly the same information.  The choice is determined by
the ease and method for finding solutions.  For example, it is
typically easier to work with the differential form when one needs to
find numerical solutions, but if one seeks an analytic solution, the
integral equations can be easier to use.  For example, in finding the
motion of a body in a potential field, one may solve the equations of
motion, or one may use the integral condition, conservation of
energy.

We argue that the integral equations provide an easier route to
deriving a unified condition for explosions.  Because the shock is a
crucial component of the core-collapse problem, when we derive the
explosion condition we embark on a route that is similar to deriving
the Rankine-Hugoniot shock jump conditions.  The route to deriving
these jump conditions involves using the integral equations in the
context of a moving boundary, the shock.

\subsection{The generic integral condition}
\label{sec:genericcondition}

As in deriving the Rankine-Hugoniot conditions, we start with the
conservation equations and the Reynolds transport theorem (or
Leibniz-Reynolds transport theorem).  Consider a
generic conserved quantity, $q$, which could represent a mass density,
momentum density, energy density, etc.  Generically, one may define
the total conserved quantity in region $D(t)$ as $\int_{D(t)}q \, dV$
and the time-rate-of-change of this conserved quantity is
$\frac{d}{dt} \left ( \int_{D(t)} q \, dV \right )$.  The Reynolds
transport theorem decomposes this time-rate-of-change into an Eulerian component and a component that accounts for the motion of the boundaries of region $D(t)$:
\begin{equation}
\label{eq:rtt}
\frac{d}{dt}\left ( \int_{D(t)} q \, dV \right ) = \int_{D(t)} \frac{\partial
  q}{\partial t}\, dV + \oint_{\partial D(t)} q {\bf v}_b \cdot d{\bf
  S}\, ,
\end{equation}
with $\partial D(t)$ the surface of this domain, $d{\bf S}$ the
surface element, and ${\bf v}_b$ the velocity of the surface.  To
make use of the Reynolds transport theorem, we need an expression for
the Eulerian time derivative, $\partial q/\partial t$.  The
conservation equation provides such an expression.  A general
conservation equation has the following form:
\begin{equation}
\label{eq:conservation}
\frac{\partial q}{\partial t} + \nabla \cdot {\bf f} = s\, ,
\end{equation}
where ${\bf f}$ is the flux of $q$, and $s$ is a source/sink of $q$.  With these two equations (eqs.~\ref{eq:rtt}~\&~\ref{eq:conservation})
we may now proceed to derive the integral conditions.

To construct the appropriate boundary value problem, we must choose the
appropriate boundaries.  We consider a boundary value problem where
the lower boundary is at $r_1$ and the upper boundary is at $r_2$.
The overall problem is divided into two domains, with a different
solution in each domain. The pre-shock solution is in the domain
$(r_s(t),r_2]$, the post-shock solution is in the domain
$[r_1,r_s(t))$, and the two solutions are joined by the Rankine
Hugoniot jump conditions at $r_s(t)$.  With the boundaries defined,
we now define the integral equations.

First, we integrate eq.~(\ref{eq:conservation}) from $r_1$ to $r_2$,
and since the boundaries are constant in time, we have
\begin{equation}
\label{eq:rightintegral}
\frac{d}{dt} \left ( \int^{r_2}_{r_1} q \, dV \right ) 
= \int^{r_2}_{r_1} \frac{\partial q}{\partial t} \, dV
= \int^{r_2}_{r_1} \left ( -\nabla \cdot {\bf f} + s \right ) \, dV\, .
\end{equation}
Next, we rewrite the left-hand side of eq.~(\ref{eq:rightintegral}) to
explicitly express the shock radius.  In order to do this, we must
divide the integral into two regions, above and below the shock, and
we use the Reynolds transport equation (eq.~\ref{eq:rtt}).  This gives
\begin{multline}
\label{eq:fullintegral}
\frac{d}{dt} \left ( \int^{r_2}_{r_1} q\, dV \right ) =
v_s 4 \pi r_s^2 ( q_{s-\epsilon} - q_{s+\epsilon}) \\
+ \int^{r_s(t)}_{r_1} \frac{\partial q}{\partial t} \, dV
+ \int^{r_2}_{r_s(t)} \frac{\partial q}{\partial t} \, dV \, ,
\end{multline}
where $s-\epsilon$ and $s+\epsilon$ represent values just below and
above the shock.  From here on, we introduce a more compact
nomenclature for the values below and above the shock with $q_-$ and $q_+$.

In the core-collapse problem, the pre-shock solution is nearly in
free-fall and can be solved analytically.  Therefore, the problem
reduces to finding the
post-shock solution subject to the lower boundary, the surface of the
NS, and the upper boundary, the shock.  Now, we take the
limit that $r_2$ approaches $r_s(t)$.  In this limit, we have the
following simplifications:
\begin{equation}
\label{eq:simplifications}
\lim_{r_2 \to r_s(t)} q_2 = q_+ \; {\rm and}
\; \lim_{r_2 \to r_s(t)} 
\int^{r_2}_{r_s(t)} \frac{\partial q}{\partial t} \, dV = 0 \, ,
\end{equation}
which, when inserted into eq.~(\ref{eq:fullintegral}), results in the simpler integral equation
\begin{equation}
\label{eq:leftintegral}
\frac{d}{dt} \left ( \int^{r_2}_{r_1} q\, dV \right ) =
v_s 4 \pi r_s^2 ( q_{-} - q_+)
+ \int^{r_s(t)}_{r_1} \frac{\partial q}{\partial t} \, dV \, .
\end{equation}
By plugging eq.~(\ref{eq:leftintegral}) into
eq.~(\ref{eq:rightintegral}), we arrive at the general integral equation
\begin{equation}
\label{eq:generalintegral}
v_s 4 \pi r_s^2 ( q_{-} - q_+)
+ \int^{r_s(t)}_{r_1} \frac{\partial q}{\partial t} \, dV
= 4 \pi (f_1 r_1^2 - f_+ r_s^2)
+ \int^{r_s}_{r_1} s \, dV \, .
\end{equation}
Note that if we take the limit that $r_1$ approaches $r_s$ from below,
this reduces to the shock jump conditions, but if we retain $r_1$ at
the neutron star surface, we have the general integral equations
describing the post-shock structure.  If we assume a steady 
profile of $q(r)$ between the neutron star, $r_1$, and the shock, $r_s$, then this reduces to
\begin{equation}
\label{eq:steadyintegral}
v_s 4 \pi r_s^2 ( q_{-} - q_+)
= 4 \pi (f_1 r_1^2 - f_+ r_s^2)
+ \int^{r_2}_{r_1} s \, dV \, .
\end{equation}

Eq.~(\ref{eq:steadyintegral}) represents the integral condition that
relates the shock velocity to the steady-state integrals.  For the
core-collapse problem, there is not one constraint, but three that
come from the three equations of hydrodynamics.  In principal, this
could be a messy algebraic problem to relate $v_s$ to the integral
conditions.  However, in the next section, we show that, in the context of the core-collapse problem, a simple relation emerges between the shock velocity and the integral conditions.

\subsection{Deriving an expression for $v_s$ in terms of the integral
  conditions}
\label{sec:derivevs}

Now, to derive the simple relation for the shock velocity
  in terms of the integral condition.  We begin by summarizing the
  traditional way of deriving the shock velocity in terms of the
  density jump and pressure jump.\footnote{The jump condition
    in the energy equation closes the system by providing an
    expression relating the jump in density with the jump in
    pressure.}  Then, we use a similar technique,
  but instead of the pressure jump, we consider the full momentum
  integral.  The Rankine-Hugoniot jump conditions in mass and momentum 
  are
\begin{equation}
\label{eq:hugoniotmass}
\rho_- u_- = \rho_+ u_+
\end{equation}
and
\begin{equation}
\label{eq:hugoniotmomentum}
P_- + \rho_-u_-^2 = P_+ + \rho_+u_+^2 \, ,
\end{equation}
where once again $\rho$ is the density and $P$ is the
  pressure.  This time, however, $u = v - v_s$ represents the velocity in the shock
  frame, which, pragmatically, is just the difference in the lab frame
  velocity ($v$) and the shock velocity.  As before, $+$
  denotes the pre-shock state and $-$ denotes the post-shock state.  Combining these two equations
  leads to the following expression for the shock velocity:
\begin{equation}
\label{eq:vshugoniot}
{\rm v}_s = \frac{v_s}{{\rm v}_+} = 
-1 + \sqrt{\frac{\beta (P_- - P_+)}{(\beta -
    1) \rho_+ v_+^2}}
\, ,
\end{equation}
where ${\rm v}_s$ is the dimensionless shock velocity, scaled by the
pre-shock speed, ${\rm v}_+ = -v_+$, and $\beta = \rho_-/\rho_+$ is
the shock compression ratio.  In deriving eq.~(\ref{eq:vshugoniot}),
there are formally two solutions, one with $+$ and one with $-$ in
front of the radical sign.  One of these solutions is real and the
other is unphysical.  The solution associated with $-$ corresponds to
$P_- - P_+ < 0$ when $\rho_- - \rho_+ > 0$; this would result in a
discontinuous rarefaction wave, which is unphysical.  The other
solution with $+$ in front of the radical sign corresponds to $P_- -
P_+ > 0$ when $\rho_- - \rho_+ > 0$; this is a shock, the only
physical solution.

So far, our derivation of eq.~(\ref{eq:vshugoniot}) is merely a summary
and does not incorporate the integral conditions;
now we use a similar procedure to derive an integral condition for the shock
velocity.  In deriving eq.~(\ref{eq:vshugoniot}), we began with the mass
and momentum jump conditions across the shock; to derive the integral
equivalent, we begin with the mass and momentum integral equations for
the shock velocity.  Using the generic integral equation,
  eq.~(\ref{eq:steadyintegral}), and the mass equation,
  eq.~(\ref{eq:mass}), the integral equation for mass that includes
  the shock is
\begin{equation}
\label{eq:masssteadyintegral}
v_s r_s^2( \rho_- - \rho_+) = \rho_1 v_1 r_1^2 - \rho_+ v_+ r_s^2 \,
\end{equation}
where the subscript $1$ denotes the state at the lower
  boundary, which in the core collapse case is the NS surface or neutrino sphere.  The
  momentum equation, eq.~(\ref{eq:momentum}), in combination with the
  generic integral equation, eq.~(\ref{eq:steadyintegral}), yields
\begin{multline}
\label{eq:momentumsteadyintegral}
v_s r_s^2( \rho_-v_- - \rho_+v_+) = 
(P_1 + \rho_1 v_1^2) r_1^2 
- (P_+ + \rho_+ v_+^2) r_s^2 \\
+ \int^{r_s}_{r_1} 2Pr\, dr - \int^{r_s}_{r_1}
GM\rho \, dr \, .
\end{multline}

In deriving these equations, we have made our first major assumption;
the solutions between the neutron star and the shock are steady.  This
implies, for example, that $\rho_1 v_1 r_1^2 = \rho_- v_- r_s^2$, but
this does not necessarily imply that $\rho_- v_- = \rho_+ v_+$.  For
the sake of brevity, we express the right-hand side of
Eq.~(\ref{eq:momentumsteadyintegral}) as $\widetilde{\Psi}$.
\begin{multline}
\label{eq:tildepsidef}
\widetilde{\Psi} = (P_1 + \rho_1 v_1^2) r_1^2 
- (P_+ + \rho_+ v_+^2) r_s^2 \\
+ \int^{r_s}_{r_1} 2Pr\, dr - \int^{r_s}_{r_1}
GM\rho \, dr \, .
\end{multline}
In the
limit of low velocities behind the shock, $\widetilde{\Psi}$ is simply
the overpressure behind the shock compared to hydrostatic
equilibrium.  The utility of using $\widetilde{\Psi}$ is that it
incorporates the entire integral solution, including the boundary
conditions.

The left-hand side of Eq.~(\ref{eq:momentumsteadyintegral}) must also satisfy 
the momentum jump condition, implying that
\begin{equation}
\label{eq:}
\widetilde{\Psi} = (P_- + \rho_-v_-^2 - P_+ - \rho_+v_+^2)r_s^2 \, .
\end{equation}
Substitution of this equation into eq.~(\ref{eq:vshugoniot}) leads to
the following dimensionless equation for the shock velocity
\begin{equation}
\label{eq:vsfullintegral}
{\rm v}_s = -1 + \sqrt{
\frac{\beta - \left ( \frac{\rho_- v_-}{\rho_+ v_+} \right )^2}
{\beta - 1}
+ \frac{\widetilde{\Psi} \beta}{(\beta - 1) \rho_+ v_+^2 r_s^2}
} \, ,
\end{equation}
where $\beta$ is the compression of density across the shock.
While formally, this expression is correct within our
  given assumptions, its relative
  complexity hides its utility.  With some simple assumptions, we
  produce a much simpler expression between ${\rm v}_s$ and
  $\widetilde{\Psi}$.  First, note that in the strong shock limit, the
ratio of densities, $\beta$, does not change as much as the jump in
pressure, so that the variation of $\widetilde{\Psi}$ dominates the
variation in ${\rm v}_s$.  If the dynamics of the shock were merely
dominated by a gamma-law EOS with $\gamma = 4/3$, then the shock compression would be
$\beta = 7$.  However, photodissociation of Fe and He nuclei at the
shock alters the energetics as material passes through the shock.  This loss of thermal energy causes a
large shock compression.  For more details on this, see \citet{fernandez09}, for
example.  As a result, the shock compression in most core-collapse
situations is $\beta \sim 9$.  With such a high shock compression,
terms such as $\beta - 1 \sim \beta$, which further reduces the
complexity in eq~(\ref{eq:vsfullintegral}).  Therefore, with fairly reasonable approximations,
eq.~(\ref{eq:vsfullintegral}) reduces to this more illuminating
expression
\begin{equation}
\label{eq:vsintegral}
{\rm v}_s \approx -1 + \sqrt{1 + \Psi} \, ,
\end{equation}
where $\Psi = \widetilde{\Psi}/(\rho_+ v_+^2 r_s^2)$ is the dimensionless overpressure normalized
  by the pre-shock ram pressure $\rho_+ v_+^2r_s^2$.  In this form, it is
clear that when $\Psi > 0$, the shock velocity is greater than zero,
${\rm v}_s > 0$.

\section{Validating the Steady-State Assumption and $\Psi \geq 0$ with one-dimensional
  parameterized simulations}
\label{sec:validate}

Later, we use $\Psi$ to propose an explosion diagnostic
for core-collapse simulations, but first we verify in this section that
$\Psi = 0$ during the stalled-shock phase and that
$\Psi > 0$ during explosion.  To perform these
validating tests, we calculate $\Psi$ in one-dimensional parameterized simulations.

The one-dimensional simulations are calculated using the new code Cufe, which will be
described fully in J.~W.~Murphy \& E.~Bloor (in preparation).  Cufe solves the equations
of hydrodynamics eqs.~(\ref{eq:mass}-\ref{eq:energy}) using higher-order Godunov
techniques.  The grid is logically Cartesian, but employs a generalized
metric allowing for a variety of mesh geometries.  For this study, we
use spherical coordinates.  The progenitor model is the 12 M$_{\odot}$ model of
\citet{woosley07}, the EOS includes effects of
dense nucleons around and above nuclear densities,\footnote{We used the
SFHo relativistic meanfield variant.} nuclear statistical
equilibrium, electrons, positrons, and photons \citep{hempel12}.  For
the neutrino heating, cooling, and electron capture we use the approximate local
descriptions of \citet{janka01} and \citet{murphy13}.

\begin{figure}[t]
  \epsscltwo
  \plotone{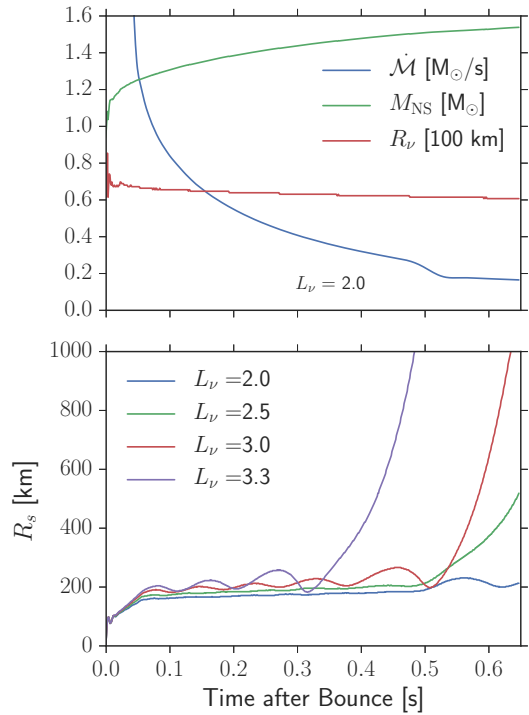}
\caption{Characteristics of one-dimensional parameterized simulations.  Top
    panel: mass-accretion rate ($\mdot$), mass of neutron star
  ($M_{\rm NS}$), and neutrino sphere or radius of neutron star
  ($\rnu$) vs. time for the
  lowest luminosity simulation.  Later we compare the integral
  condition for explosion with parameterized simulations.  Here, we
  show the three important parameters of the steady-state problem
  highlighted in Figure~\ref{fig:boundaryvalue}, which are calculated self-consistently in
  the simulations.  The evolution of these parameters with time is
  similar for all simulations, therefore we only show the lowest luminosity
  simulation here. Bottom panel: shock radius ($R_s$) vs. time for four one-dimensional parameterized
    simulations.  To facilitate comparison with the integral condition,
    we parameterize the neutrino heating and cooling.  The neutrino
    luminosities are in units of $10^{52}$ erg s$^{-1}$.  \label{fig:sim_params}}
\epssclone
\end{figure}

We recall that there are five important parameters that describe the
steady-state accretion solutions: $L_{\nu}$, $T_{\nu}$, $M_{\rm NS}$,
$\rnu$, and $\mdot$.  Two of these, $L_{\nu}$ and
$T_{\nu}$, we set in the parameterized simulations; the rest we
calculate self-consistently given the equations of hydrodynamics, the
progenitor structure, and the EOS.
The top panel of Figure~\ref{fig:sim_params} shows the time evolution of $M_{\rm
  NS}$, $\rnu$, and $\mdot$; we highlight the post-bounce phase
of a model that does not explode.  $\mdot$ starts quite high, well above 1
M$_{\odot}$/s, but then drops down to $\sim$ 0.2 M$_{\odot}$/s near
the end of the simulation.  At around 500 ms after bounce, $\mdot$ has
a significant drop, which is a result of a density shelf advecting through the
shock.  Throughout this paper, you will notice that this significant
drop in $\mdot$ becomes imprinted on many of the explosion diagnostics,
including the shock radius vs. time in the bottom panel of Figure~\ref{fig:sim_params}.

When compared to more realistic simulations \citep[see Figure 4]{melson2015}, $\rnu$
  evolves very little during the time shown in
  Figure~\ref{fig:sim_params}.  This is a known drawback of the
  standard light-bulb prescription, but it does not impact the
  qualitative conclusions that we present later.  In more realistic
  calculations such as \citet{melson2015}, the proto-neutron star contracts for two
  reasons.  For one, as matter piles onto the PNS, the neutron star compresses
  a little.  The primary reason that the proto-neutron star
  contracts, however, is that the core neutrino luminosity cools the neutron
  star.  In our neutrino model, we omit the diffusive cooling of
  neutron star.  Hence, our core only contracts as a result of the added
  weight of matter.

However, because each moment may be modeled as a successive
  set of steady-state solutions, this lack of contraction does not
  affect our general conclusion that we may calculate an explodability
  parameter.  Later, we will calculate an
  explodability parameter that depends upon the five parameters,
  $\rnu$ being one of them.  Because the solutions are
  time independent, the explodability parameter calculation is also
  time independent.  Each moment in time has its own explodability
  value that is independent of any other moment.  Therefore, it does
  not matter what the evolutionary history of $\rnu$ is; we are able
  to calculate the explodability parameter for any neutrino-sphere
  radius history.

To sample a range of explosion timescales, we vary the light-bulb
neutrino luminosity, $L_{\nu}$.  The bottom panel of
Figure~\ref{fig:sim_params} shows the resulting shock radii vs. time
labeled by $L_{\nu}$ in units of $10^{52}$ erg s$^{-1}$.  In all models, the
shock forms at 153 ms after the start of the simulations and quickly
stalls between $\sim$150 to $\sim$200 km.  The general trend is that
the higher luminosity models explode earlier than lower luminosity
models.  The lowest luminosity model does not explode at all.
However, it does experience a significant outward adjustment and
oscillation of the shock as $\mdot$ drops significantly around 500 ms
after bounce.  The $L_{\nu} = 2.5$ \& 3.0 models explode during the
advection of the density shelf.  An obvious feature present in the
shock radius evolution plot are the shock radius oscillations.  They
are often present in one-dimensional simulations near explosion
\citep{ohnishi06,buras06a,murphy08b}, and \citet{fernandez12} suggests
that they might be related to the advective-acoustic feedback loop
that is responsible for the standing accretion shock instability (SASI) in multi-dimensional simulations.

\begin{figure*}[t]
  \epssclone
  \plotone{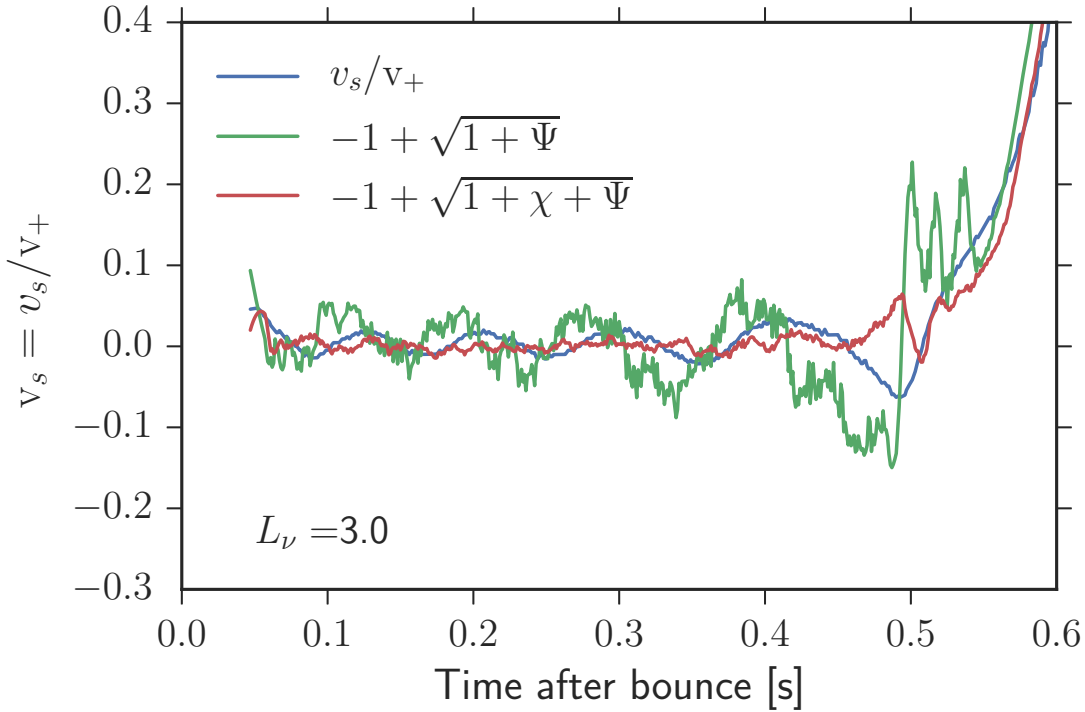}
\caption{Normalized shock velocity vs. time after bounce.  This plot
  validates the assumptions and derivation of
  eq.~(\ref{eq:vsintegral}), which relates the shock velocity
${\rm v}_s$ to pressure integral condition $\Psi$.  ${\rm v}_s$ is the 
  shock velocity normalized by the the pre-shock in-fall speed, 
${\rm v}_+$, and $\Psi$ is overpressure behind the shock in integral
  form.  When $\Psi \approx 0$, ${\rm v}_s \approx 0$, 
and when $\Psi > 0$, ${\rm v}_s > 0$.  This validation sets the stage for the explodability condition in Figure~\ref{fig:nearness}. \label{fig:vsobs_psitheory}}
\epssclone
\end{figure*}

Our first task is to verify one of the primary results of
  this manuscript, the relationship between ${\rm v}_s$ and
  the integral quantity $\Psi$, eq.~(\ref{eq:vsintegral}).
  Figure~\ref{fig:vsobs_psitheory} plots the shock velocity normalized
  by the pre-shock in-fall speed, ${\rm v}_s = v_s/{\rm v}_+$, for the
  $L_{\nu} = 3.0$ model.  For comparison, we plot the right-hand side
  of eq.~(\ref{eq:vsintegral}), $-1 + \sqrt{1 + \Psi}$, where $\Psi$
  is calculated directly from the simulation.  For the most
  part, the curves agree, validating our derivation and assumptions.

Next, we validate the steady-state assumption.  Before we do so, we
need to consider how time dependence might affect
eq.~(\ref{eq:vsintegral}).  In steady-state, the momentum integral
equation is $v_s r_s^2 (\rho_-v_- - \rho_+v_+) = \widetilde{\Psi}$,
but if we include the time-dependent term, then the momentum integral equation is
\begin{equation}
\label{eq:momintegraldt}
v_s r_s^2 (\rho_-v_- - \rho_+v_+) = \widetilde{\chi} +
\widetilde{\Psi} \, ,
\end{equation}
where 
\begin{equation}
\widetilde{\chi} = \int_{r_1}^{r_s - \epsilon} \frac{\partial (\rho
  v)}{\partial t} \, dV \, ,
\end{equation}
and $r_1$ is the size of the proto-neutron star, or
  neutrino-sphere radius.
Our goal is to see how this time-derivative term affects our expression
for the shock velocity, therefore we use eq.~(\ref{eq:momintegraldt}) in our
derivation for the shock velocity in section~\ref{sec:derivevs}.
The equivalent of the final result, eq.~(\ref{eq:vsintegral}), is
\begin{equation}
\label{eq:vsintegraldt}
{\rm v}_s \approx -1 + \sqrt{1 + \Psi + \chi} \, ,
\end{equation}
where $\chi$ is just $\widetilde{\chi}$ normalized by $\rho_+v_+^2r_s^2$.

\begin{figure}[t]
  \epssclone
  \plotone{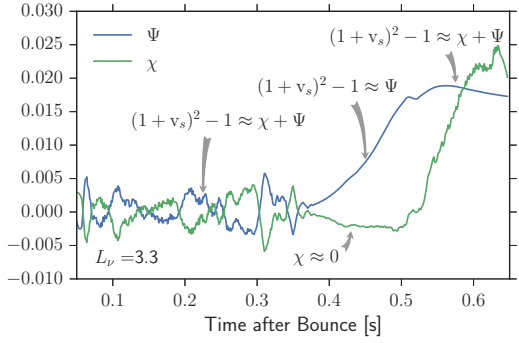}
\caption{Checking that steady-state is a reasonable assumption in the
  integral condition.  Here we show the overpressure, $\Psi$ and
  time-derivative of momentum, $\chi$, as a function of time after
  bounce.  During the stalled-shock phase, the average shock velocity
  is very nearly zero, and $\Psi$ and $\chi$ oscillate about zero as
  the shock oscillates.  Even with these oscillations, both $\Psi$ and
  $\chi$ are quite small.  During the initial stages of explosion,
  $\chi \approx 0$, and the amplitude of the shock velocity is best
  represented by $\Psi$ alone. Later, as the explosion proceeds vigorously, both
  $\Psi$ and $\chi$ contribute to the velocity.  Since we are
  concerned with the revival phase of the shock, we find that
  steady-state is a reasonable approximation during initiation of the
  explosion. \label{fig:cond_sim_psi_dpdt}}
\epssclone
\end{figure}

In some sense,
Figure~(\ref{fig:cond_sim_psi_dpdt}) has validated
Equation~(\ref{eq:vsintegral}) in which we ignored the time-derivative
term, so that it already hinted that $\chi$ would be small.  To be certain
that $\chi \approx 0$, we compare $\Psi$ and $\chi$ in
Figure~(\ref{fig:cond_sim_psi_dpdt}).  In general, 
$(1+{\rm v}_s)^2 - 1 = \Psi + \chi$, but if $\chi \approx 0$, then we
expect $(1+{\rm v}_s)^2 - 1 \approx \Psi$.  During the steady-state
phase, when the average shock velocity is zero, we find that both
$\Psi$ and $\chi$ are small and approximately zero.  In fact, their
small amplitudes nearly cancel.  During the initiation of the explosion, $\Psi$ grows
substantially, but $\chi \approx 0$.  Later, when the
explosion really takes off, both $\chi$ and $\Psi$ are important.
Interestingly, we find that assuming steady-state is a valid
approximation during the steady-state phase and the initiation of the
explosion, but once explosion commences in earnest, one must
clearly consider a time-dependent evolution.  The focus of this
manuscript is in deriving a condition for the initiation of explosion,
however, in which case Figure~\ref{fig:vsobs_psitheory} suggests that
eq.~(\ref{eq:vsintegral}) is a fine place to start.

\begin{figure*}[t]
  \epssclone
  \plotone{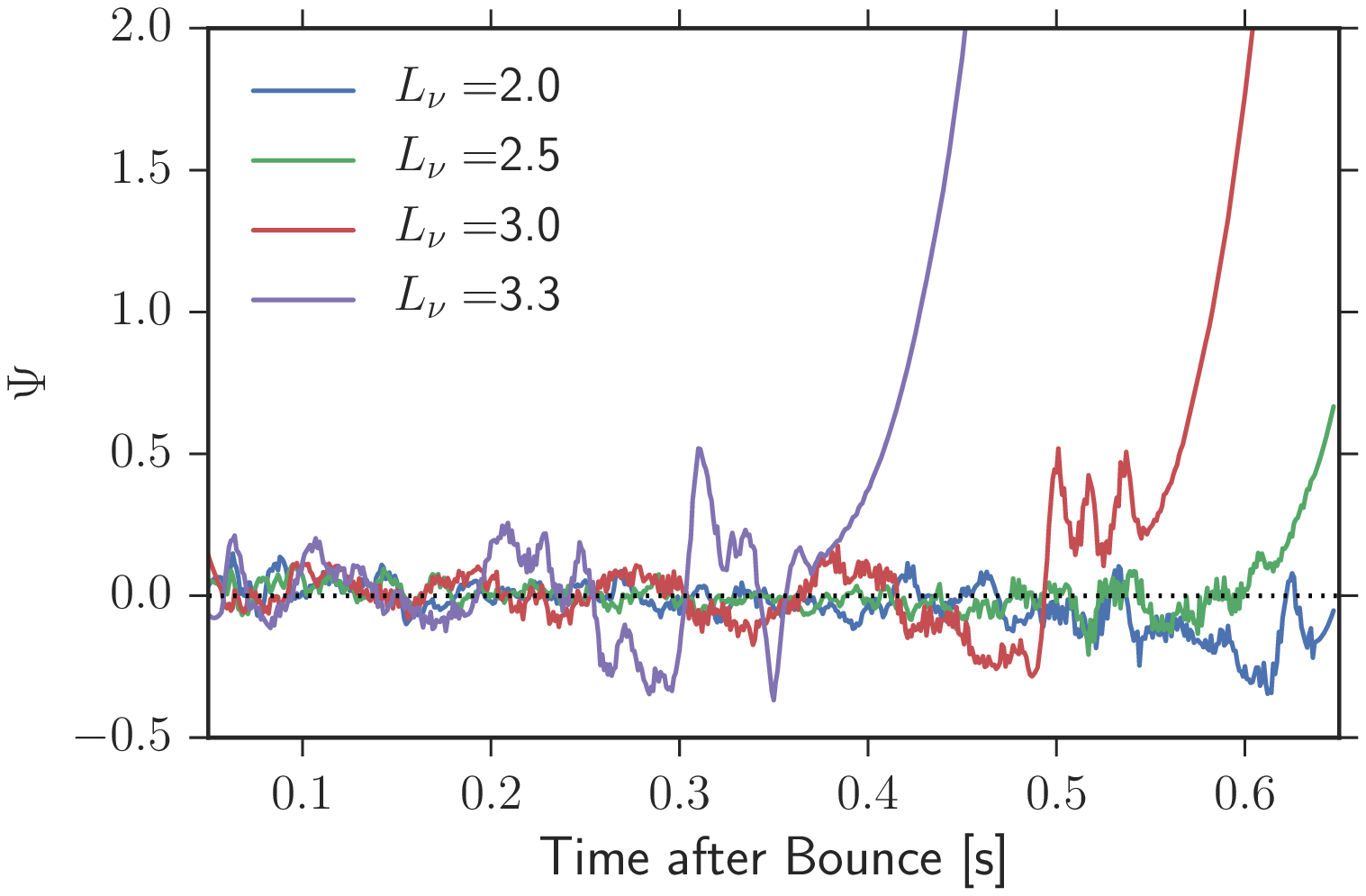}
\caption{$\Psi$, the integral explosion condition, evaluated from the one-dimensional
  simulations as a function of time.  In section~\ref{sec:integralcondition}, we derive an
  integral condition for explosion based upon the momentum equation;
  we derive that $v_s \geq 0$ corresponds to $\Psi \geq 0$.  To test
  whether this condition corresponds to explosion, we
  evaluate $\Psi$ for the one-dimensional parameterized simulations described in
  section~\ref{sec:validate}; we find that $\Psi = 0$ during the stalled-shock
  phase, and $\Psi > 0$ during explosion.  These results strongly suggest that
  $\Psi \geq 0$ is a useful condition for explosion.  However, this
  condition does not by itself predict explosion.  It only indicates
  that explosion has occurred.  In section~\ref{sec:explodability}, we propose a method to
  turn $\Psi \geq 0$ into a useful explosion diagnostic. \label{fig:cond_sim_psi}}
\epssclone
\end{figure*}

In Figure~\ref{fig:cond_sim_psi}, we show $\Psi$ for all
  of the models that we considered. $\Psi$ is indeed zero during the
steady-state phase and $\Psi > 0$ during explosion.
The fact that simulations roughly validate $\Psi \geq 0$
implies that $\Psi$ is a good dimensionless explosion diagnostic.

\section{$\psimin$: A nearness-to-explosion condition}
\label{sec:explodability}

The fact that $\Psi >0$ during explosion
(Figure~\ref{fig:cond_sim_psi}) suggests that $\Psi$ has the potential
to be a good explosion
diagnostic.  However, just calculating $\Psi$ from the
simulations is not the most useful condition; $\Psi$ remains
near zero during the
non-exploding phase, and it only deviates from zero while the simulation is
exploding.  This behavior is not a very useful explosion diagnostic,
and one might as well only use the shock radius.  In fact, most
explosion ``conditions'' or ``diagnostics'' to date have this problem.
They do not provide a nearness-to-explosion condition.  What we need
is a way to translate the $\Psi \geq 0$ condition into a useful
nearness-to-explosion condition.  Fortunately, $\Psi$ lends itself to
such an explosion diagnostic.

\begin{figure*}[t]
  \epssclone
\plotone{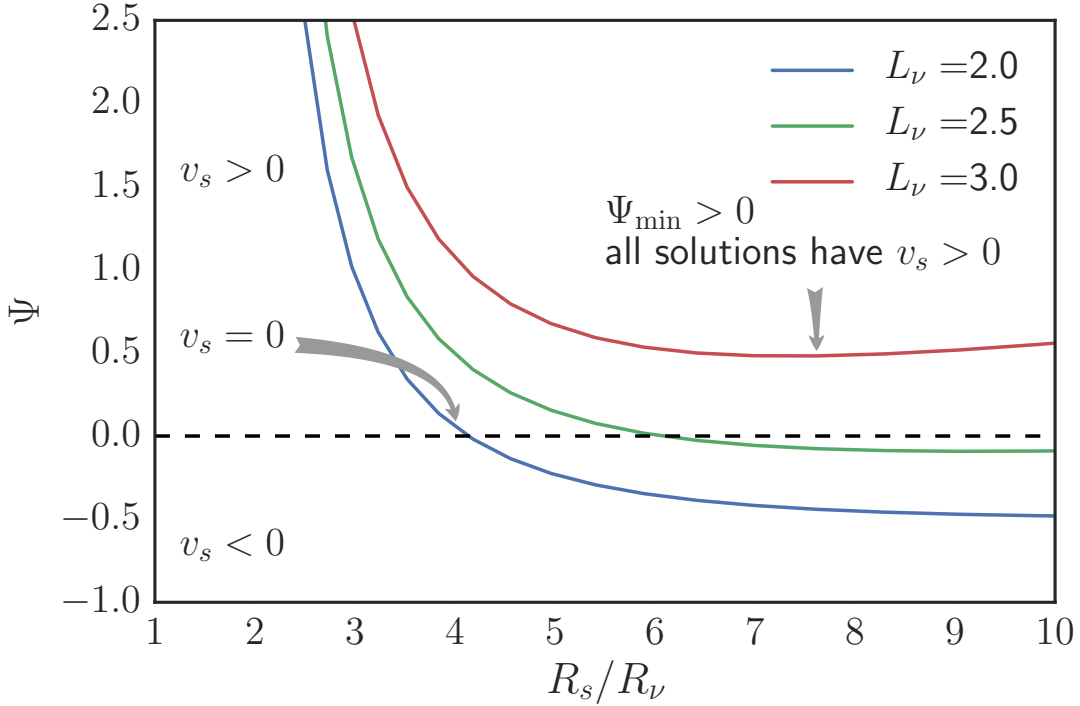}
\caption{The family of steady-state solutions represented by their
  values of $\Psi(R_s/\rnu)$ for three different neutrino luminosities but all else being equal.   For some parameters, $\Psi$ may
  be negative, zero, or positive.  The solution with
  $\Psi = 0$ ($v_s = 0$) corresponds to the equilibrium solution.  For
  each family, there is a solution with a minimum $\Psi$, which we
  denote $\psimin$.  When $\psimin < 0$, a stalled-shock ($v_s = 0$)
  solution exists, but when $\psimin > 0$, only solutions with
  $v_s > 0$ exist.  We propose that $\psimin > 0$ is an excellent
  explosion diagnostic (see Figure~\ref{fig:nearness}). \label{fig:psi_xs_lum}}
\epssclone
\end{figure*}

\begin{figure}[t]
  \epsscltwo
\plotone{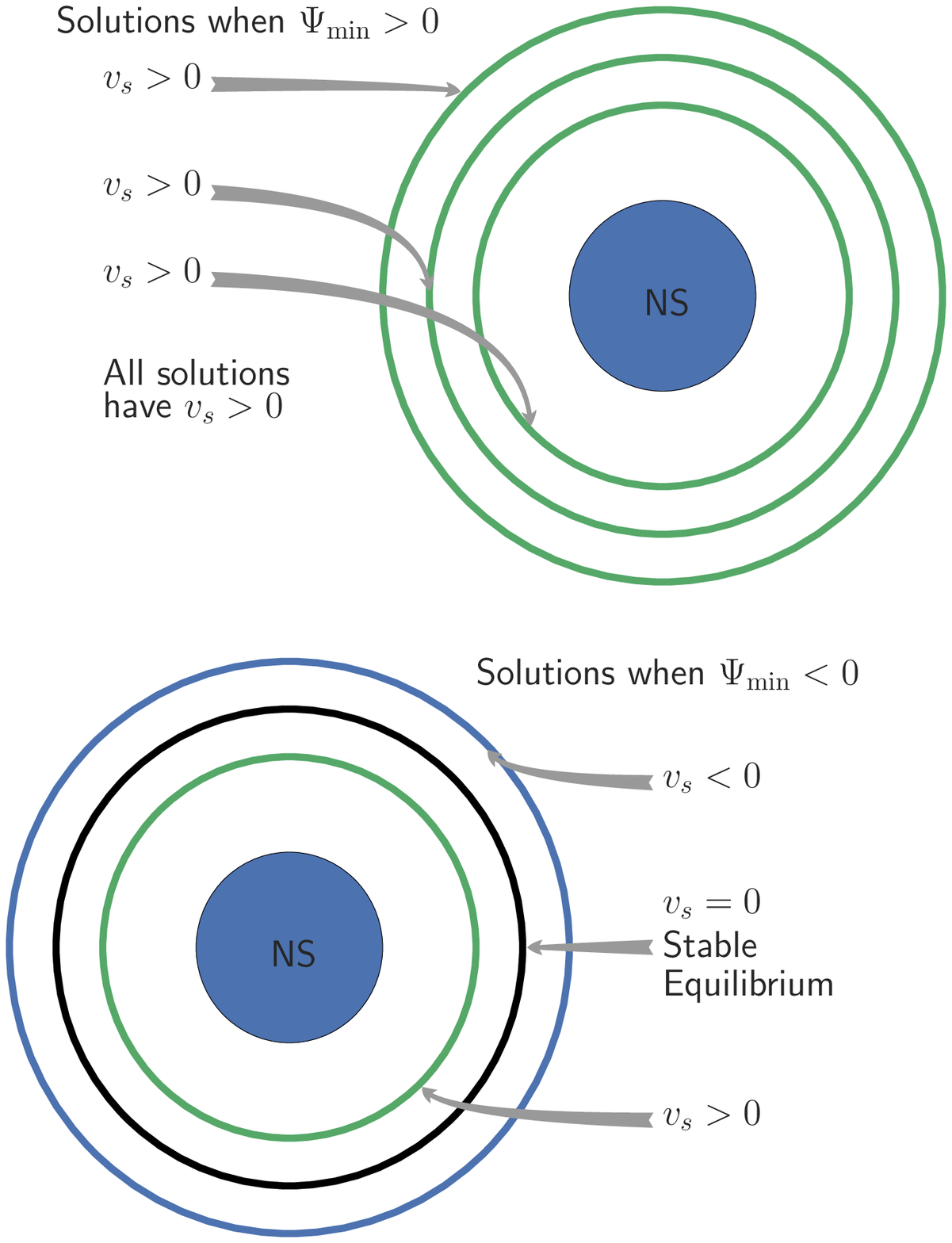}
\caption{This is an illustration to demonstrate the two regimes seen
  in Figure~\ref{fig:psi_xs_lum}.  If $\psimin < 0$, then there are solutions
  with $v_s < 0$, $v_s = 0$, and $v_s > 0$.  In general, the solution
  with $v_s = 0$ is an equilibrium solution, and as long as it
  exists, simulations (and presumably Nature) settle on this
  steady-state stalled-shock solution.  When $\psimin > 0$,
  the only solutions are those in which $v_s > 0$.  Therefore,
  $\psimin$ seems to be a natural condition for explosion.  \label{fig:sol_diag}}
\end{figure}

Our strategy for developing a nearness-to-explosion condition is to
extract $L_{\nu}$, $T_{\nu}$, $\rnu$, $M_{\rm NS}$, and $\mdot$
from the simulations and calculate the quasi-steady-state solutions for
this set of parameters.  For a given set of parameters, there is a family
of solutions to the quasi-steady-state equations, each with a
different shock radius.
\footnote{Far from the equilibrium shock radius, these solutions are
  likely incorrect in detail since they correspond to situations that
  are far from a steady-state.  However, our approach does not require
  accurate models far from equilibrium}
Each solution has a value for $\Psi$
which may be $< 0$, $= 0$, or $> 0$, which corresponds to solutions
with $v_s < 0$, $v_s = 0$, or $v_s > 0$, respectively.  The $v_s = 0$ ($\Psi = 0$) solution is a quasi-equilibrium
solution.  We will show that for each family of solutions, there is always a minimum
$\Psi$, which we denote $\Psi_{\rm min}$.  When $\psimin < 0$,
then $v_s=0$ solutions exist.  However, when $\psimin > 0$,
the only solutions that exist have $v_s > 0$.  Therefore, we propose
that $\psimin$ is an excellent explosion diagnostic providing a
nearness-to-explosion condition.  Figures~\ref{fig:psi_xs_lum}~\&~\ref{fig:sol_diag} illustrate these
points.

Having outlined the strategy, we now provide the specifics in
calculating $\psimin$.  We begin with the steady-state equations
\begin{equation}
\label{eq:steadymass}
\dot{M} = 4 \pi r^2 \rho v \, ,
\end{equation}
\begin{equation}
\label{eq:steadymom}
\rho \frac{d v}{d r} + \frac{d P}{d r} = - \rho \frac{\phi}{r} \,
\end{equation}
and
\begin{equation}
\label{eq:steadyene}
\frac{\dot{M}}{4 \pi r^2} \frac{d \varepsilon}{dr} + P \frac{d(r^2
  v)}{r^2 dr} = \rho q \, .
\end{equation}
To solve these equations, we need an EOS to relate P in
terms of $\varepsilon$ and $\rho$, and we need boundary conditions.
For the region that we consider, above the neutrino sphere and below the
shock, the dominant contributors to the EOS are neutrons, protons,
helium nuclei ($\alpha$), positrons, electrons, and photons.
Because this region is not dominated by dense nuclear physics,
the EOS is more straightforward and for the most part analytic.
Eventually, we hope to translate the results of this manuscript into
an analytic solution for the conditions for explosion.  To facilitate
that later goal, we use the analytic EOS now.  We treat
the neutrons, protons, and $\alpha$s as an ideal nonrelativistic gas;
 the positrons, electrons, and photons constitute the relativistic
 part of the plasma, and we consider the positrons and electrons in an
 arbitrary degeneracy.  To calculate the relative abundances of
 neutrons, protons, and $\alpha$s, we assume nuclear statistical
 equilibrium for these three components and use the Saha equation to
 calculate their abundances.  See appendix~\ref{app:eos} for the
 details in calculating the EOS and the abundances.

In the spirit of \citet{burrows93} and others \citep{yamasaki05,yamasaki06,yamasaki08}, we solve the steady-state
equations, Eqs.~(\ref{eq:steadymass}-\ref{eq:steadyene}), using the
Rankine-Hugoniot jump conditions for a stalled shock.  The jump
conditions give the post-shock state (subscript $-$) in terms of the
pre-shock state (subscript $+$):
\begin{equation}
\rho_- v_- = \rho_+ v_+ \, ,
\end{equation}
\begin{equation}
\rho_- v_-^2 + P_- = \rho_+ v_+^2 + P_+ \, ,
\end{equation}
and
\begin{equation}
h_- + \frac{v_-^2}{2} = h_+ + \frac{v_+^2}{2} \, .
\end{equation}
For the pre-shock state, we make the following assumptions that enable
analytic solutions of the conditions.  First, we assume that $\mdot = 4
\pi r_s^2 \rho_+ v_+$ is
a constant.  Second, we assume that the star is accreting onto the
shock at a large fraction of free fall, so the pre-shock speed is given by 
$v_+^2 = 2 \alpha_v \phi (R_s)$, where $\phi(R_s)$ is the
  potential at the shock radius, and $\alpha_v$ is the fraction of free-fall.  Third, we assume that the pre-shock
Bernoulli constant is roughly zero, 
$h_+ + v_+^2 - \phi(R_s) \approx 0$.
Finally, we assume a gamma-law relationship for the pre-shock
pressure, internal energy, and density:
$\gamma_4 \equiv P/(\rho \varepsilon) + 1$.
Under these assumptions, the pre-shock pressure is
\begin{equation}
P_+ = \frac{\gamma_4 \rho_2 (1 - \alpha_v)}{\gamma_4 - 1}\phi(R_s) \, .
\end{equation}
Specifying $\mdot$, $\alpha_v$, $\gamma_4$, and $R_s$ then sets the
boundary conditions at the shock.  When \citet{burrows93} first introduced the
critical luminosity, they specified one more boundary condition at the base:
$\tau = \int \kappa \rho dr = 2/3$.  This extra condition enables one
to self-consistently solve for the shock radius that permits a
steady-state-stalled-shock solution.  \citet{yamasaki05} noted that the density at
the radius where $\tau = 2/3$ is almost always the same, so that one can
easily replace the $\tau$ condition by specifying a specific density, $\rho_{2/3}$,
at the neutrino sphere, or neutron star surface.  The traditional way
to find the critical curve is to look for solutions that satisfy
$\rho_1 = \rho_{2/3}$; above a curve in
luminosity and accretion-rate space there are no solutions that
satisfy this condition.  \citet{burrows93} interpreted this as a
critical condition for explosion.  However, it has never been
  clear why this condition on the inner density should naturally lead
  to a critical condition for explosion.  Now, we show that the solutions
above the curve indeed only have solutions that correspond to 
${\rm v}_s > 0$.

To do this, we
  show that a condition on the inner boundary, 
$\rho_{2/3}/\rho_1 > 1$, is equivalent to $\Psi > 0$ and ${\rm v}_s >
  0$.  Historically, the method for finding the critical curve is as
  follows.  First, for a given $R_s$, one finds ``solutions'' to the
  steady-state equations by starting with the $v_s = 0$ jump conditions at
  the shock and integrating steady-state equations inward to the inner boundary.  The
  resultant partial solution satisfies the governing equations except that
  they do not necessarily satisfy the inner boundary condition, which
  means that they are not a
  true solution.  One then finds the true solution by modifying the
  shock radius until the partial solution also gives the correct inner
  boundary condition.  One may use these partial solutions at any shock radius
  to infer what $v_s$ would be as a function of $R_s$.

Understanding the density profile and pressure profiles of these
partial solutions is the key to connecting the old condition on the
inner density with a condition on $v_s$.  Note that $\Psi$ in eq.~\ref{eq:vsintegral} 
depends upon $P_1$, $\rho$, and $\rho_1$.  If one could specify
analytic solutions for the pressure and density profile, then the
condition would be analytic.  In a future paper (J.~W.~Murphy \&
J.~C.~Dolence, in preparation), we will do just
that.  For now, we use the numerical solutions to the steady-state
equations to propose a semi-analytic solution for the explosion
condition.  It is our experience that the density profile of the
partial solution is similar for a
given set of parameters and $R_s$.  In other words, the density
profile may be written as $\rho(r) = \rho_1 z(r)$.  Furthermore, a
natural dimensionless expression relating the density and pressure is
$y = P/(\rho \phi)$, and we find that the shape and scale of $y$
mostly depends on the five parameters and $R_s$.  Therefore, we find
that $P$ is also proportional to $\rho_1$ via 
$P = \rho_1 y(r) z(r) \phi(r)$.  Substituting these expressions for
$\rho$ and $P$ into the expression for $\widetilde{\Psi}$ (Equation~\ref{eq:tildepsidef}),
one finds that
\begin{equation}
\label{eq:psirho1}
\widetilde{\Psi} = \rho_1 f - P_+r_s^2 - \rho_+ v_+^2 r_s^2 \, ,
\end{equation}
where $f$ incorporates $y(r)$, $z(r)$, and $\phi(r)$ in the first,
third, and fourth terms of eq.~(\ref{eq:tildepsidef}).  When one integrates the steady-state
equations, one naturally finds that $\widetilde{\Psi} = 0$ but the
inner density $\rho_1$ is not necessarily equal to the desired inner
boundary, $\rho_{2/3}$.  Because $\rho = \rho_1 z(r)$, then we may
easily substitute $\rho_{2/3}$ to see how $\widetilde{\Psi}$ changes when we demand that
$\rho_1 = \rho_{2/3}$.  In terms of $\rho_1$, the inner density of the
partial solution, and $\rho_{2/3}$, the desired inner boundary, the
dimensionless overpressure becomes
\begin{equation}
\label{eq:psi_in_terms_of_rho1}
\frac{\widetilde{\Psi}}{(P_++\rho_+v_+^2)r_s^2} =
\frac{\rho_{2/3}}{\rho_1} - 1 \approx \Psi \, ,
\end{equation}
where the final approximation comes about because the
  pre-shock pressure is typically much lower than the ram pressure,
  $P_+ \ll \rho_+ v_+^2$.  By eq.~(\ref{eq:vsintegral}), this approximate expression
  for $\Psi$ gives the following expression for the shock velocity in
  terms of the inner densities:
\begin{equation}
{\rm v}_s \approx -1 + \sqrt{\frac{\rho_{2/3}}{\rho_1}} \, .
\end{equation}

This presents our final derivation to show that the solutions above
the critical luminosity-accretion-rate curve have positive shock
velocity.  Before this derivation, \citet{burrows93} and many others suggested
that the inability to find solutions that have $\rho_1 = \rho_{2/3}$
implies explosion.  In fact, our derivation implies the following
explosion condition:
\begin{equation}
\frac{\rho_{2/3}}{\rho_1} > 1 \rightarrow \Psi > 0 \rightarrow {\rm
  v}_s > 0 \, .
\end{equation}

To find $\Psi$ as a function of $R_s$, we use the one-dimensional simulations to
inform the values for $\mdot$, $M_{\rm NS}$, $\rnu$, $\alpha_v$, and $\gamma_4$, and
of course, we set $L_{\nu}$ and $T_{\nu}$ to the values used for the
one-dimensional parameterized simulations.  Then we find solutions to the
steady-state equations for a wide range of shock radii.  These
solutions represent a family of solutions at different $R_s$ but all
with the same parameters.  For each solution and associated $R_s$, we
evaluate $\Psi$ using eq.~(\ref{eq:psi_in_terms_of_rho1}), where we follow \citet{yamasaki05} and
set $\rho_{2/3} = 7 \times 10^{10}$ g cc$^{-1}$.  Since ${\rm v}_s \approx -1
+ \sqrt{1 + \Psi}$, $\Psi$ naturally shows which solutions have $v_s < 0$, $v_s
= 0$, or $v_s > 0$.  The $v_s = 0$ solution corresponds to the
steady-state-stalled-shock solution.

Figure~\ref{fig:psi_xs_lum} shows the outcome of this process.  Each line corresponds to a specific set of parameters and
one family of solutions.  For clarity, we only vary $L_{\nu}$ between different families of solutions.  For low values of
$L_{\nu}$, all three solutions ($v_s > 0$, $v_s = 0$, and
$v_s < 0$) are possible.  For low $R_s$, $\Psi > 0$, implying that $v_s > 0$ and
the shock would move outward.  For larger 
$R_s$, $\Psi < 0$ so that $v_s < 0$
and the shock would tend to move inward.  In between, at a very
specific shock radius, there is a solution that has 
$\Psi = 0$ and $v_s = 0$.  This solution represents an equilibrium
solution.   While solutions far from equilibrium are not steady and
therefore are poorly represented by the steady-state equations, our
approach does not rely on the quantitative accuracy of this
representation.  For a visual aide to these concepts, see
Figure~\ref{fig:sol_diag}.

Note that for high values of $L_{\nu}$ there are no solutions for which $\psimin < 0$.  In these cases, all solutions correspond to $v_s > 0$.
There is a very specific set of parameters for which $\psimin
= 0$.  The locus of such parameters defines a critical hypersurface that generalizes but encompasses the
critical neutrino luminosity condition of \citet{burrows93} (see Section~\ref{sec:comparison} for
more on this).

\begin{figure*}[t]
  \epssclone
\plotone{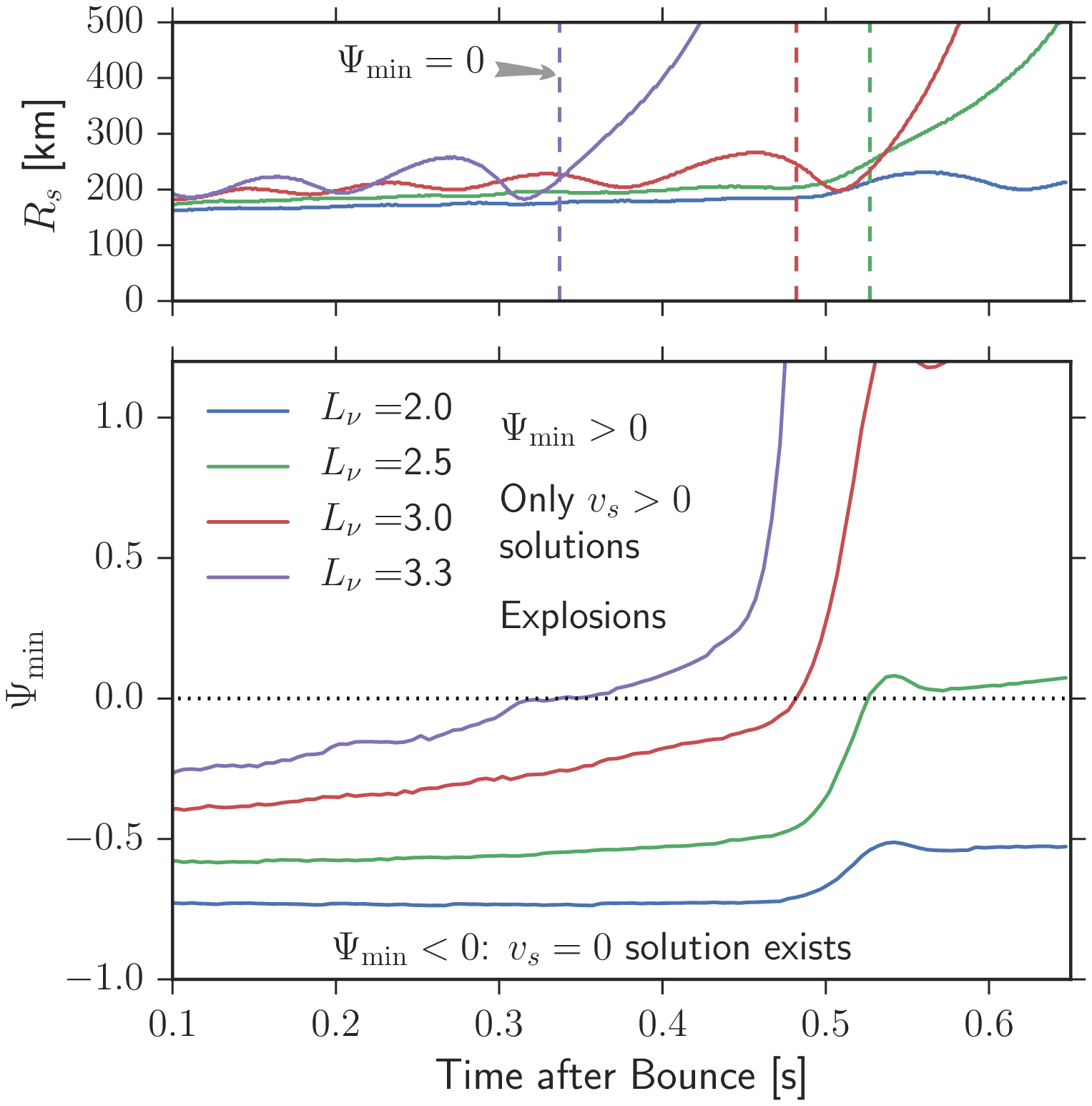}
\caption{\label{fig:nearness} Nearness-to-explosion condition for one-dimensional
  hydrodynamic simulations:
  $\psimin$ vs. time.  For each time, we extract the important
  parameters of the problem ($L_{\nu}$,
  $T_{\nu}$, $\rnu$, $M_{\rm NS}$, and $\mdot$) from the one-dimensional
  hydrodynamic simulations and calculate
  the family of steady-state solutions.  $\psimin$ represents the minimum
  possible value of $\Psi$ over that family (see Figure~\ref{fig:psi_xs_lum}).  While $\psimin < 0$,
  there exists a stalled accretion shock solution ($\Psi = 0$ and $v_s
  = 0$).  When $\psimin > 0$, all possible solutions have $v_s > 0$, which we associate
with explosion.  Empirically, we find that before explosion $\psimin <
0$ and during explosion $\psimin > 0$.  Most importantly, the value of
$\psimin$ offers an excellent indication of nearness-to-explosion.}
\epssclone
\end{figure*}

To validate that $\psimin$ satisfies the desired qualities of an
explosion diagnostic, we plot $\psimin$ in Figure~\ref{fig:nearness} for the one-dimensional parameterized simulations.  The parameters that we set by
hand in each simulation are $L_{\nu}$ and $T_{\nu}$; we keep $T_{\nu}$
set at 4 MeV, and vary $L_{\nu}$.  At each time, we extract the other important
parameters $M_{\rm NS}$, $\rnu$, and $\mdot$, which are
calculated self-consistently in the simulation.  We then calculate the
family of steady-state solutions for that set of parameters and find
the value of $\psimin$.  Figure~\ref{fig:nearness} shows that $\psimin$ has the
desired qualities of an explosion condition.  Before explosion, $\psimin
< 0$ and during explosion, $\psimin > 0$.  Moreover, we suggest that the value of
$\psimin$ before explosion provides a useful metric for how far away
the simulation is from explosion.

\section{Comparison with Other Explosion Conditions}
\label{sec:comparison}

In this section, we briefly compare the integral explosion diagnostic,
$\psimin$ with three other conditions: the critical neutrino luminosity
condition \citep{burrows93}, a timescale ratio condition
\citep{janka98,thompson00,thompson05,buras06b,murphy08b}, and the antesonic
condition \citep{pejcha12}.  In
this manuscript, we mostly compare the effectiveness of these
conditions with the integral condition; in a forthcoming paper (J.~W.~Murphy \&
J.~C.~Dolence, in preparation), we
show that one can actually derive all three conditions from the
integral condition by making various further approximations.  For now, we simply compare the conditions.

\begin{figure*}[t]
  \epssclone
\plotone{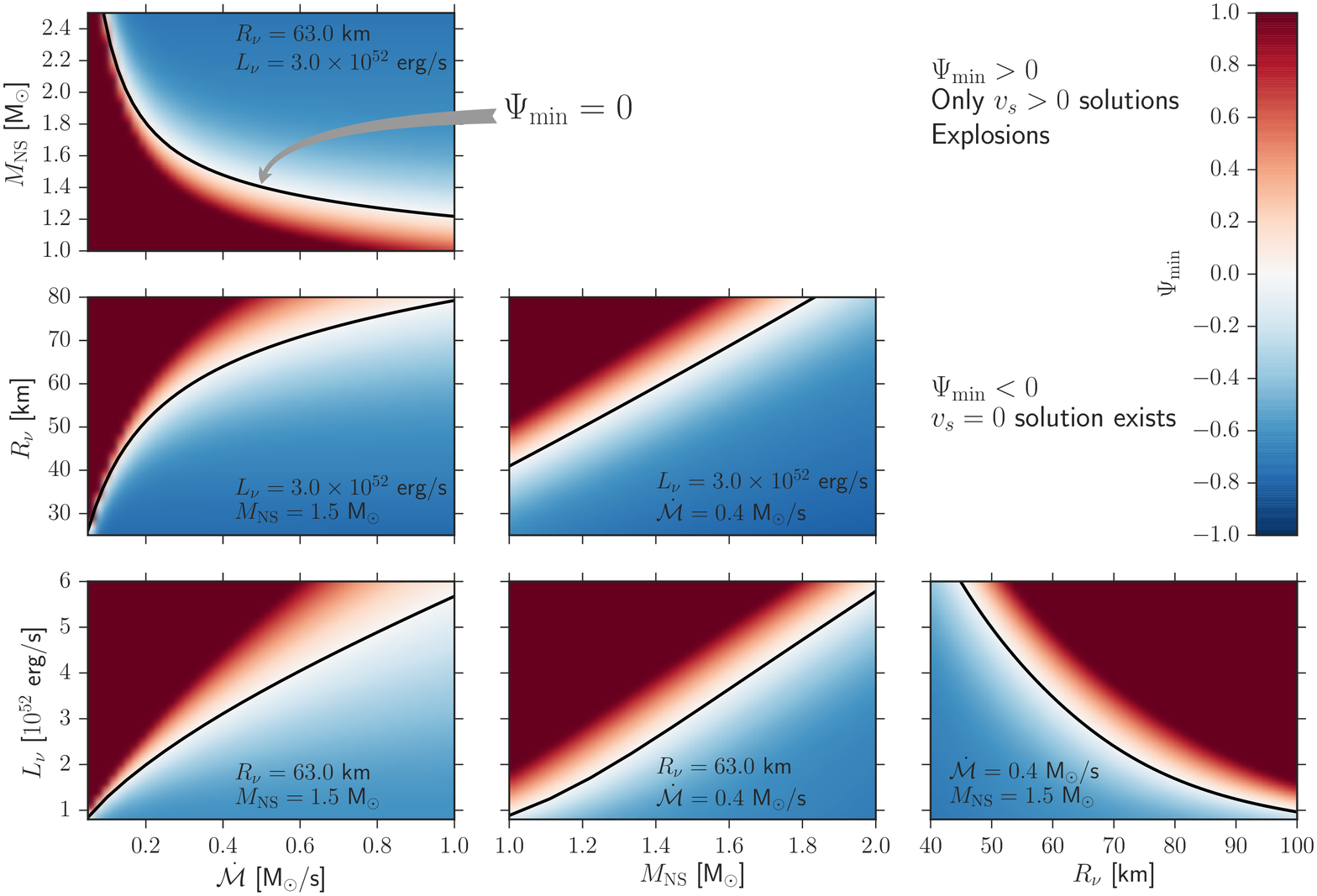}
\caption{ ``Critical curves'' defined by the $\psimin = 0$ condition.
  These plots show that the $\psimin = 0$ is another way to define the
  critical curve, the solutions above the critical curve have
  positive shock velocity, and the $L_{\nu}$-$\mdot$ curve is merely a
  slice of a critical hypersurface.  Each panel represents a two-dimensional slice through the five-dimensional
  parameter space.  In all
  panels, $T_{\nu}$ is fixed to 4 MeV.  The two other fixed parameters
  are labeled in each panel.  The color map shows the
  value of $\psimin$ as a function of the two remaining parameters,
  and the solid black line shows where $\psimin = 0$. In the lower
  left panel, we show that the $\psimin = 0$ condition produces the
  critical $L_{\nu}-\mdot$ curve of \citet{burrows93}.  Note that
  above this curve $\psimin > 0$, which implies that solutions above
  the curve have $v_s > 0$. In the context of our
  assumptions, this suggests that the solutions above the critical
  curve are explosive, as \citet{burrows93} initially suggested.
  Finally, with these panels, we show that the critical neutrino
  luminosity curve is a slice of a more general critical
  hypersurface.  Most importantly, this critical hypersurface is
  represented by one dimensionless condition, $\psimin = 0$. \label{fig:critcurves}}
\end{figure*}

Of these three, the most closely related condition is the critical
luminosity condition of \citet{burrows93}.  In fact, a primary motivation in
deriving the integral condition is to derive a condition that shows
that the solutions above the critical neutrino luminosity curve
correspond to $v_s > 0$.  In section~\ref{sec:integralcondition}, we derived the integral
condition for $v_s > 0$, and now, in the bottom left panel of Figure~\ref{fig:critcurves}, we show
that this same integral condition reproduces the critical neutrino
luminosity curve of \citet{burrows93}.

To reproduce the critical neutrino luminosity curve in the lower
  left panel of
Figure~\ref{fig:critcurves}, we first fix three of the five important parameters
of the problem, $\rnu$, $M_{\rm NS}$, and $T_{\nu}$.  In other
words, we restrict the dimensionality of the integral condition to the
$L_{\nu}$-$\mdot$ plane.  Then we
find the solutions to the steady-sate equations (eqs.~\ref{eq:steadymass}-\ref{eq:steadyene}) and select
the solutions in the $L_{\nu}$-$\mdot$ plane that have
$\psimin = 0$.  The locus of these specific solutions forms the solid curve
that we show in Figure~\ref{fig:critcurves}.  Above this curve, 
$\psimin > 0$ and therefore all solutions have $v_s > 0$.  Below this curve,
$\psimin < 0$, and therefore a steady-state stalled-shock
solution exists.  This curve is exactly the same critical neutrino
luminosity curve that \citet{burrows93} and others have derived.  The
difference is in the way that it is derived.  \citet{burrows93} solved the
steady-state equations looking for solutions for which the shock and
$\tau = 2/3$ conditions are satisfied.  There is no statement about
the behavior of $v_s$ above the curve.
Rather than highlighting $\tau$, we instead use the $\psimin = 0$
condition, that one readily sees that $v_s > 0$ above the curve.

In addition to showing that $v_s > 0$ above the critical neutrino luminosity
curve, we also show that $\psimin > 0$ is a more general explosion
condition than the critical $L_{\nu}$-$\mdot$ curve.  In fact, in the
important parameters, $\psimin = 0$ represents a critical hypersurface,
and the critical $L_{\nu}$-$\mdot$ curve is just one slice of
this more general critical condition.  To help illustrate this,
  we use in Figure~\ref{fig:critcurves}
the $\psimin = 0$ to derive other critical curves, where each is just
a different slice of the critical hypersurface.  Each
critical curve is constructed in the same way as the
$L_{\nu}$-$\mdot$ critical curve, except that a different set of three
parameters is fixed.  In all panels, $T_{\nu}$ is fixed to 4 MeV;
each panel shows the values for the other two parameters.
In the past, the literature has focused on the $L_{\nu}$-$\mdot$ critical curve, mostly
because that is how this useful critical curve was discovered and
presented \citep{burrows93}.  However, Figure~\ref{fig:critcurves}
clearly shows that there are many other
critical curves, one for each pair of the five parameters.

By focusing on one critical curve, one runs the
danger of overemphasizing two of the parameters and underemphasizing
the other three parameters.  For example, the $L_{\nu}$-–$\mdot$ critical
curve suggests that one should pay attention to the evolution of $L_{\nu}$
and $\mdot$.  When it comes to the equations, however, there is
nothing special about $L_{\nu}$ or $\mdot$; all parameters are potentially
important.  Depending upon the evolution of the parameters, it might
therefore be more instructive to consider the evolution of, for
example, $R_{\nu}$ and $\mdot$.

The $\psimin=0$ critical condition suggests another way to view
  the nearness-to-explosion that does not overemphasize any one
  parameter.  The five parameters define a five-dimensional hyperspace,
  and the $\psimin=0$ condition defines a hypersurface within the
  five-dimensional space.  Below the $\psimin=0$ hypersurface,
  steady-state solutions exist, and above this hypersurface we show
  that the solutions are likely explosive.  As a simulation evolves,
  it will trace out a path in this 5-dimensional hyperspace, and in
  general, it will move toward or away from the $\psimin=0$ hypersurface.
  Therefore, a generalized distance to this surface would be an excellent
measure of nearness-to-explosion.  In a future paper, we will
present discussions of this generalized distance. For now, we merely
introduce this new concept, and propose that since the value of
$\psimin$ defines the hypersurface, it is a
good proxy for the generalized distance to the bounding
hypersurface. This definition is useful for two reasons.  One, since
the hypersurface is defined by $\psimin=0$, we are able to
define the critical condition by one dimensionless parameter and not
five.  Two, this dimensionless parameter is intimately related to a
fundamental question of CCSN theory.  $\Psi$ is
proportional to $v_s$, and a fundamental question of CCSN theory is
``what are the conditions for $v_s > 0$?''  Hence a single
dimensionless parameter, $\psimin$, that defines the condition for explosion is
also intimately related to an important question in CCSN theory.

\begin{figure}[t]
  \epsscltwo
\plotone{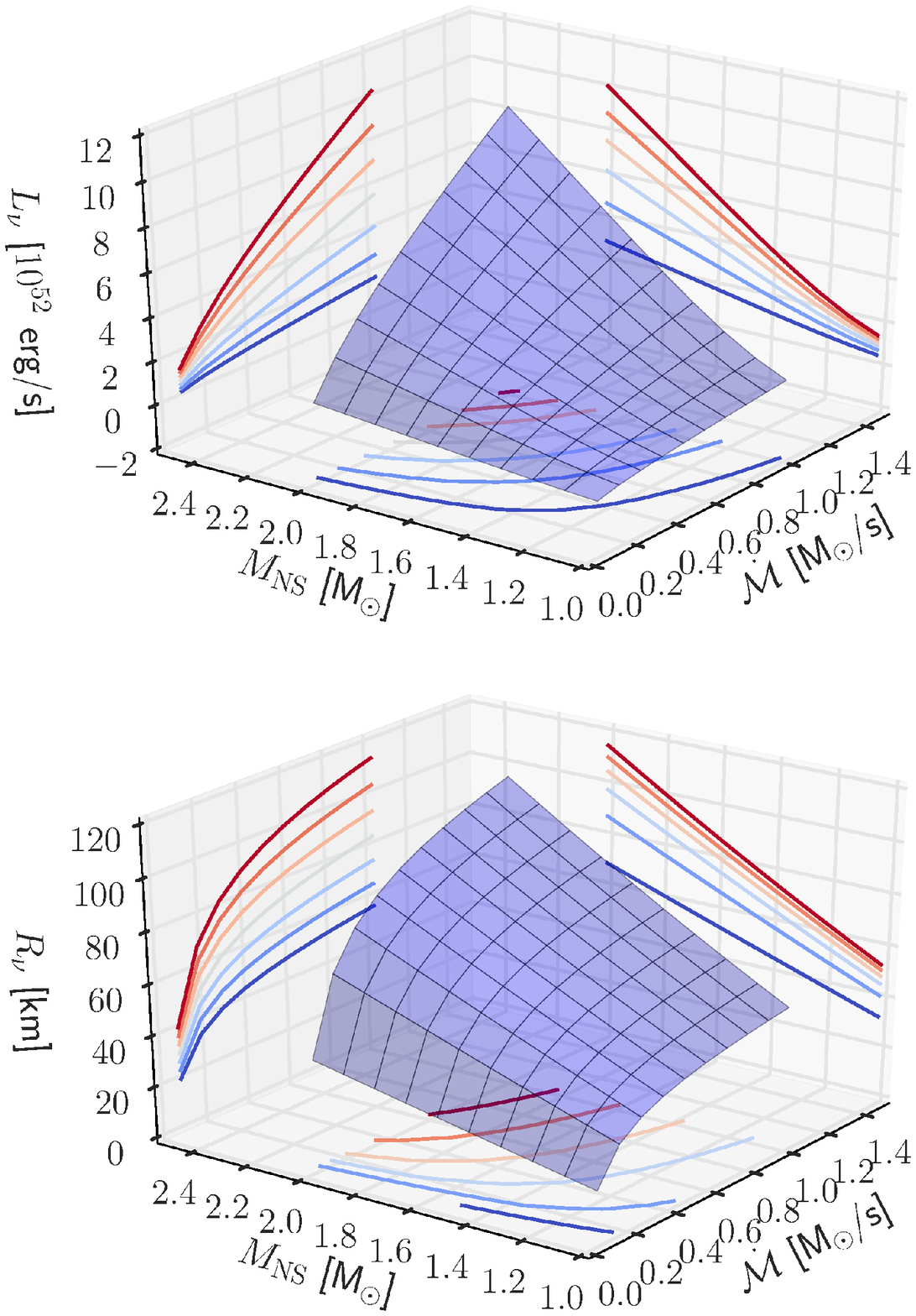}
\caption{Critical hypersurface for explosion.  The $\psimin = 0$
    condition defines a hypersurface in the five-dimensional parameter
    space.  Below the hypersurface, $\psimin < 0$, and steady-state
    solutions exist; above the hypersurface, $\psimin > 0$.
    Therefore, only solutions with $v_s > 0$ exist.  It is difficult to
    visualize the full hypersurface; therefore we present two
    slices that are themselves two-dimensional surfaces. In both plots, we fixed
    $T_{\nu}$ to 4 MeV.  In the top plot, we also fixed $\rnu$ to 63
    km, and in the bottom plot we fixed $L_{\nu}$ to $3 \times
    10^{52}$ erg s$^{-1}$.  We also show ``critical curves,'' for which we
    specify the value for the third
    dimension. \label{fig:3dprojections}}
\epssclone
\end{figure}

To help illustrate the critical hypersurface,
Figure~\ref{fig:3dprojections} shows two slices of this hypersurface.
Since it is difficult to visualize a five-dimensional hypersurface, we
instead show two slices of the hypersurface.  In the top panel of
Figure~\ref{fig:3dprojections}, we fix $T_{\nu}$ to 4 MeV and $\rnu =
63$ km.  The resulting surface is a two-dimensional surface defined by
three dimensions.  In each of the axis planes, we also show critical
curves, with each curve corresponding to a specific value of the third axis.  The bottom panel shows an equivalent slice,
but this time the slices are at $T_{\nu} = 4$ MeV and 
$L_{\nu} = 3 \times 10^{52}$ erg/s.  While it is impossible to show the
full hypersurface, we hope to illustrate with these two-dimensional
slices that the critical condition is not only a critical curve, but
a critical hypersurface defined by $\psimin = 0$.  Below the
hypersurface, $\psimin < 0$, and steady-state solutions exist; above
the hypersurface, $\psimin > 0$ and all solutions likely have $v_s > 0$.

Another common class of explosion metrics are the timescale ratios in
which one compares an advection timescale ($\tau_{\rm adv}$) to a
heating timescale ($\tau_{\rm heat}$).  Roughly, neutrino heating
might be expected to significantly modify the internal energy of the
accreting material only when $\tau_{\rm heat} \lesssim \tau_{\rm adv}$.
Based on this, many have used the ratio $\tau_{\rm adv}/\tau_{\rm
  heat}$ as a diagnostic to indicate how close a model is to explosion
for a given snapshot \citep{janka98,thompson00,thompson05,buras06b,murphy08b}.  While this may seem sensible at face value, the timescale ratio
arguments all neglect important aspects of the CCSN problem ---
cooling that occurs below the gain region, for example.
Meanwhile, \citet{murphy08b} and others more recently
\citep{dolence13,takiwaki14,hanke12} have shown that the turbulent dynamics of multidimensional
lead to longer advection times on average, speculating that this may
be an important effect vis-\`a-vis 1D vs. multi-D.

There are many approximate definitions of $\tau_{\rm adv}$ and
$\tau_{\rm heat}$ \citep{janka98,murphy08b,fernandez12} .  Early definitions of the advection time include $\tau_{\rm adv} =
\int_{r_{\rm gain}}^{r_{\rm shock}} dr/v_r$, but the most recent
definitions use the mass in the gain region and the mass-accretion
rate
\begin{equation}
\label{eq:tadv}
\tau_{\rm adv} = \frac{M_{\rm gain}}{\mdot} \, .
\end{equation}
For the rest of this paper, we adopt this most recent definition.  For
the heating timescale, the generic approach is to compare a total
energy with an integrated heating rate $\tau_{\rm heat} = E/Q$.
Generically, many define the heating rate as $Q = \int_{\rm gain}
(H-C) \rho \, dV$.  The total energy has several definitions, and
since none are derived from a firm explosion condition, all are
arbitrary.  We highlight and use one condition; we consider
the total internal energy in the gain region $E = \int_{\rm gain}
\varepsilon \rho \, dV$.  The choice of both the energy scale and the
region are completely arbitrary and not derived from an actual
explosion condition.  In a forthcoming paper (J.~W.~Murphy \&
J.~C.~Dolence, in preparation), we will derive a
timescale ratio condition from the integral condition, but for now we
just compare to the arbitrary definitions from the past.  
We pick one example, the total internal energy in the
gain region,
\begin{equation}
\label{eq:theat}
\tau_{\rm heat} = \frac{E_{\rm int}}{Q} \,  .
\end{equation}

\begin{figure}[t]
  \epssclthree
\plotone{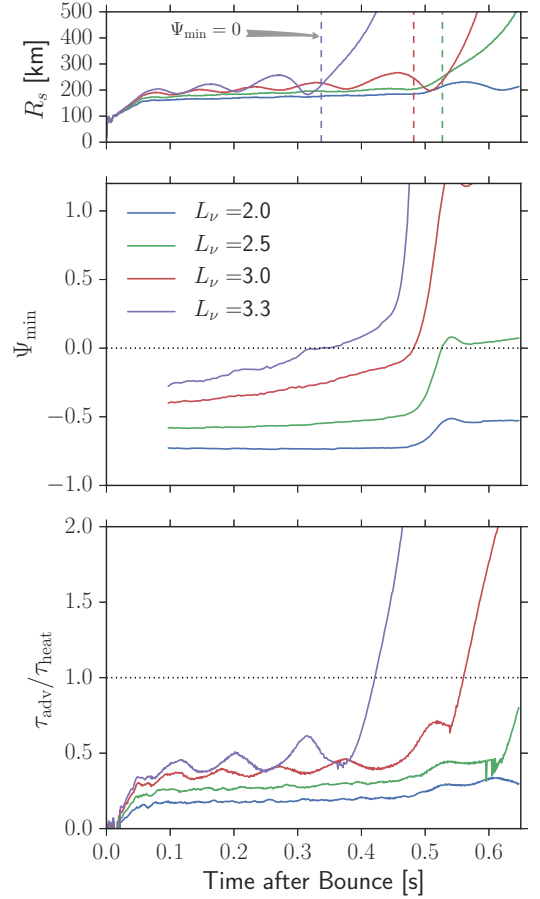}
\caption{A comparison of the explosion diagnostic, $\psimin > 0$, and a timescale ratio
  condition, $\tratio$.  The top plot shows
  $\psimin$ and the bottom plot shows the ratio of the
  advection time ($\tau_{\rm adv}$) to the heating time ($\tau_{\rm
    heat}$). The advection timescale is defined as $\tau_{\rm adv} =
  M_{\rm gain}/\mdot$, where $M_{\rm gain}$ is the mass in between the
  gain radius and the shock.  We adopt a heating timescale that is
  a ratio of energy and heating rate in the gain region:
  $\tau_{\rm heat} = E/Q_{\rm gain}$, where $Q_{\rm gain} =
  \int_{r_{\rm gain}}^{R_s} (H - C)\rho \, dV$, and $E = \int_{\rm
    gain} \varepsilon \rho \, dV$  is the total internal energy in the
  gain region.  The $\tratio$ is an approximate heuristic explosion
  diagnostic.  The choice of the terms to define the timescales and
  the region over which to integrate are arbitrary.  The timescale ratio
  is $\mathcal{O}(1)$ but it shows no sign of a reliable ``critical''
  value or utility in predicting explosion.  In fact, it is only as good
  as using the shock radius as an explosion diagnostic.  On the
  other hand, $\psimin$ works quite well as an explosion diagnostic.\label{fig:timescales}}
\end{figure}

Figure~\ref{fig:timescales} compares our derived integral condition, $\psimin$, with the approximate timescale ratio condition, $\tratio$.  As one might expect,
the rough heuristic condition does not perform as well at predicting
explosions.  Indeed, $\tratio$ is $\mathcal{O}(1)$ before explosion and
increases dramatically after explosion.  However, the ratio has
serious problems as an explosion metric.  For example, one
might be tempted to conclude that the sharp upward trend in this ratio
would be a good indicator of explosion.  However, careful inspection
of $\tratio$ and the shock radii
in Figure~\ref{fig:timescales} shows that $\tratio$ just follows $R_s$.  Therefore,
using $\tratio$ is not much more useful than $R_s$.  One might be
tempted to use the simulations to calibrate a ``critical'' value for
$\tratio$.  However, the results in the literature as well as those
shown here indicate that no such ``critical'' value
exists \citep{muller12,hanke12,dolence13}.

\begin{figure}[t]
  \epssclfour
\plotone{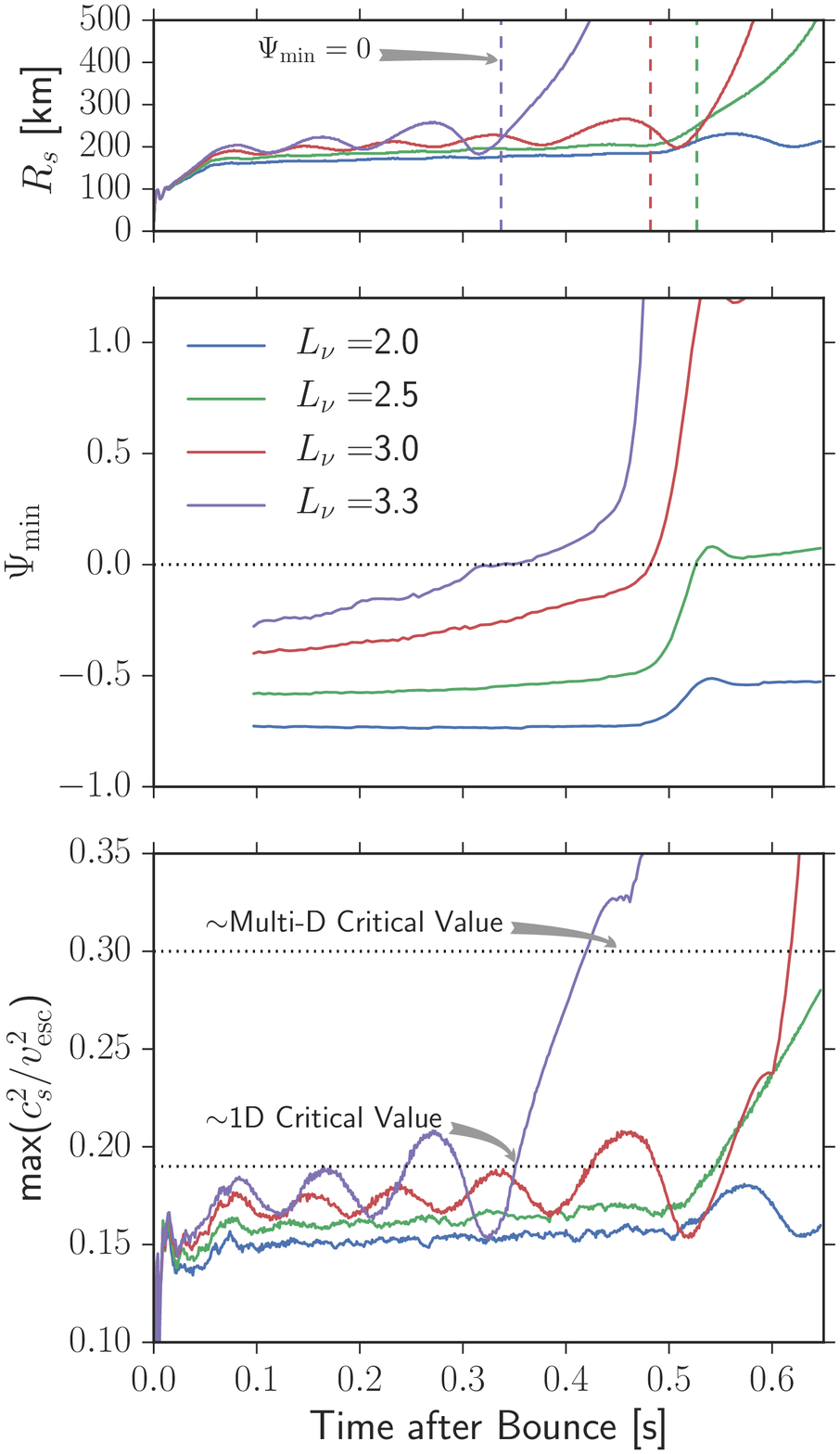}
\caption{The $\psimin$ explosion diagnostic in comparison with the
  antesonic condition of \citet{pejcha12}.  For an isothermal flow in
  one dimension, \citet{pejcha12} show that there are no
  steady-state accretion solutions that match the pre-shock flow for
  $c_T^2/v_{\rm esc}^2 > 3/16$.  By analogy, they propose that a
  similar condition exists for the nonisothermal situation of the
  core-collapse problem.  By numerically exploring the steady-state
  solutions in one-dimensional approximations,
  they propose that $\max(c_s^2/v_{\rm esc}^2) \gtrsim 0.19$ divides
  steady-state accretion from explosion.  In multidimensional simulations, \citet{mueller12}
  and \citet{dolence13} empirically find that $\max(c_s^2/v_{\rm esc}^2) \gtrsim
  0.3$ is a better ``critical'' value.  The bottom panel of this
  figure shows that the one-dimensional ``critical'' value is a decent indicator of
  explosion.  It certainly performs better than the timescale ratio
  seen in Figure~\ref{fig:timescales}.  Still, the $\psimin > 0$ condition has
  some practical advantages.  For one, $\psimin > 0$ is an
  integral condition derived from the supposition that $v_s > 0$
  corresponds to explosion.  Second, the dynamic range of $\Psi_{\rm
    min}$ illustrates more clearly which simulations are far from
  explosion.  Third, while the lowest luminosity model appears to
  always approach explosion for the $\max(c_s^2/v_{\rm esc}^2)$
  condition, the $\psimin$ condition clearly shows that the
  $L_{\nu} = 2.0$ model moves away from explosion, except when the density shelf
  accretes through the shock, but then again it moves away from
  explosion.  Fourth, $\max(c_s^2/v_{\rm esc}^2) \gtrsim 0.19$ does
  not predict explosion:  the actual antesonic condition overshoots
  the proposed critical value by 10\% before it actually explodes,
  which is half of the full dynamic range of the antesonic condition
  ($\sim 20$\%);  when the antesonic condition looks as if it indicates explosion, it does so well after the initiation of
  explosion.  \label{fig:antesonic}}
\end{figure}

The final comparison is between $\psimin$ and the antesonic
condition of \citet{pejcha12}; see Figure~\ref{fig:antesonic}.   Similar to our motivation, \citet{pejcha12}
were motivated to explain the origin of the critical
luminosity curve and possibly derive a more general condition.  They
took a slightly different tack.  \citet{pejcha12} considered the family of
accretion and wind solutions and hypothesized that explosion occurs at
the intersection between steady-state accretion solutions and
steady-state wind solutions.  By considering isothermal profiles, they
were able to define an analytic condition that divides these
solutions.  For $c_T^2/v_{\rm esc}^2 > 3/16$, only wind solutions are
possible, which they attributed to the transition between steady-state
accretion and explosion.  Given the great deal of neutrino heating and
neutrino cooling, the post-shock profile is certainly not isothermal.
Therefore, they were unable to derive a similar analytic condition in a more realistic CCSN context.
Instead, they solved the steady-state equations and empirically
searched for a similar condition that would be appropriate for the
core-collapse case.  The resulting explosion condition that they
proposed is the antesonic condition, $\max(c_s/v_{\rm esc})^2 \gtrsim 0.19$, where $\max()$
denotes the maximum in between the NS and the shock.  Empirically, our
one-dimensional simulations indicate that a value of $\sim$0.19 is about right.
In multidimensional simulations, \citet{mueller12} and \citet{burrows13} found
a much higher value of $\sim$0.3.  However, given that the antesonic
condition was derived and discussed in the context of one-dimensional spherical
symmetry, it is no surprise that the multidimensional simulations do
not match.  After all, we derive an integral condition in the same
spherically symmetric context.  In a forthcoming paper, we rederive
the integral condition including turbulence, and find that the new
integral condition is consistent with simulations.  We suspect that
one could do the same for the antesonic condition.
However, to use the antesonic
condition in practice, one must measure the ``critical'' condition in
simulations ex post facto.  While the antesonic
condition, which is analytically derived in the isothermal case and
empirically derived in the nonisothermal case, is imperfect,
Figure~\ref{fig:antesonic} shows that it fairs a little better
than the timescale ratio condition.

While the antesonic condition fares slightly better than the timescale
ratio at predicting explosion, it still remains less favorable
compared to the integral condition (middle panel of
Figure~\ref{fig:antesonic}).  For one, like the timescale ratio, the
dramatic upward trend in the antesonic ratio that might be attributed
to explosion actually occurs well after explosion; this means that one
can rule out using the upward trend as an explosion diagnostic.  One of
the useful features of the integral condition, $\psimin > 0$ is
that one can tell even before explosion that a model is unlikely to
explode.  There are two features that enable this: 1) the closeness of
$\psimin$ to zero, and 2) the trend toward or away from zero; for
example, $\psimin$ for the $L_{\nu} = 3.0$ model marches toward
explosion, while $\psimin$ for the $L_{\nu} = 2.0$ model marches away
from explosion.  Similar to point 1) for the integral condition, the
antesonic condition is indeed closer to the ``critical'' value before
explosion.  However, even the antesonic diagnostic for the $L_{\nu} =
2.0$ marches toward the ``critical'' value even though it never
explodes.  Furthermore, the antesonic diagnostic overshoots the
``critical'' value by $\sim$10\% of the ``critical'' value.  This may
seem like a triumph of estimating the ``critical'' value, but the
dynamic range of the antesonic diagnostic is only $\sim$20\% of the
``critical'' value, making the estimate only $\sim$50\% accurate when
considering the dynamic range of the antesonic diagnostic.

\section{Discussion and Conclusions}
\label{sec:conclusions}

For many years, the critical neutrino luminosity curve has been the most impactful model for understanding successful explosion
conditions.  However, no one had demonstrated analytically that the solutions above
the curve are explosive, nor had
anyone successfully used the critical neutrino luminosity as an
explosion diagnostic for core-collapse simulations.  In this
manuscript, we took steps to show that the solutions above the
  critical neutrino luminosity have $v_s > 0$, and we used this
  derivation to propose an explosion diagnostic for core-collapse simulations.  When we began
this project, our primary goal was to merely show that the solutions
above the critical neutrino luminosity curve are explosive.  We have
partially done so by deriving an integral condition, $\psimin \geq 0$,
for $v_s \geq 0$ and showing that the solutions above
the critical neutrino luminosity curve have 
$v_s > 0$.  Along the way to deriving this integral condition, we discovered that the neutrino luminosity curve is a projection of a
more general critical hypersurface.  Most strikingly, this critical
hypersurface is represented by a single
dimensionless parameter, $\psimin \geq 0$, and it promises to
be a useful explosion diagnostic for core-collapse simulations.

In deriving the integral condition $\Psi \geq 0$, we made two simple but profound approximations: (1) we considered
steady-state solutions, and (2) we considered that $v_s > 0$ corresponds to explosion.
With these two approximations, we derived an integral condition for
$v_s \geq 0$ and recast it in a dimensionless form, $\Psi \geq 0$.
We verified this integral condition using one-dimensional
parameterized simulations and found that during the stalled-shock
phase $\Psi \approx 0$ and during explosion $\Psi > 0$.  Although the
comparisons with simulations successfully verified the integral condition,
they also demonstrated that simply calculating $\Psi$ from the simulations
is not a useful explosion diagnostic.  Because the simulations always
find a stalled-shock solution before explosion, $\Psi$ is always zero
before explosion.  Such a diagnostic fails to indicate a distance to
explosion.  If, on the other hand, we extract the important parameters
from the simulations and calculate all of the possible steady-state
solutions, then a better diagnostic emerges.  

The steady-state solutions fall into one of two broad categories:
in one category, a stalled-shock solution exists, and in the other,
only $v_s > 0$ solutions exist.  To understand the origin of these
categories, one must first understand that there are five important parameters that characterize
each steady-state solution: $L_{\nu}$ (neutrino luminosity), $T_{\nu}$
(neutrino temperature), $\rnu$ (neutrino-sphere radius or proto-neutron star radius),
$M_{\rm NS}$ (proto-neutron star mass), and $\dot{M}$ (accretion
rate).  For a fixed set of parameters, there is a
family of solutions, and each solution in this family has a different
value of $\Psi$.  In general, $\Psi$ may be negative, zero, or
positive, and because $\Psi \propto v_s$, each solution may have
$v_s < 0$, $v_s = 0$, or $v_s > 0$.  If a $v_s = 0$ solutions exists, it
represents the preferred quasi-equilibrium solution.  Generically, for each
family of solutions, there is a minimum $\Psi$ that we denote
$\psimin$, and it is this parameter that determines whether a solution
belongs to the stalled-shock category or to the explosive category.  If
$\psimin < 0$, then the family of solutions has solutions with $v_s <
0$, $v_s = 0$, and $v_s > 0$ with the $v_s = 0$ being the preferred
quasi-equilibrium solution.  If on the other hand, $\psimin > 0$, then
only $v_s > 0$ solutions exist.  If we attribute $v_s > 0$ to explosions,
then $\psimin$ is a useful explosion diagnostic.

We calculated $\psimin$ for several one-dimensional parameterized simulations
(see Figure~\ref{fig:nearness}) and find that the time evolution of $\psimin$ makes an excellent
explosion diagnostic.  From the time that we track $\psimin$, one can
tell whether a simulation is near explosion.  For those that are near
explosion, $\psimin$ tends to march toward explosion throughout much
of the simulation.  The model that does not explode shows a general
evolution of $\psimin$ away from explosion.  Next, we need to explore
$\psimin$ in the context of multidimensional and self-consistent
neutrino radiation hydrodynamic simulations to see if these promising behaviors remain.

In section~\ref{sec:comparison} and Figures~\ref{fig:timescales}~\&~\ref{fig:antesonic}, we compared this new integral condition
with explosion conditions that have been defined previously.
Specifically, we compared $\psimin > 0$ to a timescale ratio
condition and the antesonic condition.  In contrast to the integral
condition, the $\tratio$ condition is heuristic, only good to order
unity, shows no obvious ``critical'' value, and is no better than simply
using $R_s$ as an explosion diagnostic.  The
antesonic condition fares better than the timescale ratio
condition.  Even so, it lacks the accuracy and predictive attributes
of the integral condition.

To derive the integral condition, we made two approximations that
must be investigated further: we assumed that $v_s > 0$ corresponds to
explosion, and we assumed steady-state throughout our derivation.  These are two very strong assumptions that could have
invalidated our derivation.  For example, $v_s > 0$ does not
necessarily equate to explosion, especially when we relax the
steady-state condition.  Oscillations of the shock are one example in
which $v_s > 0$ may manifest but not lead to explosion.  In order to
make progress, we decided to see where these approximations led.  In
the end, we were able to derive an integral explosion diagnostic that
successfully describes the explosion conditions of parameterized one-dimensional
simulations.  However, we did not show that once $\psimin > 0$ and the
shock begins to move outward that it continues to move outward.  In
some sense, one expects the shock to continue to move out, because
once $\psimin > 0$ the only steady-state solutions are solutions with $v_s
> 0$, but this does not preclude the possibility of a dynamic (non-steady-state) solution that is oscillatory.  A next step in confirming
our derivation is to show that once $\psimin > 0$, the dynamic
solution is a solution in which $v_s > 0$ continues.

We found the integral condition to be a successful explosion diagnostic
for one-dimensional parameterized explosions, but we suspect that it can do much
more.  We suspect that the integral condition may be an excellent
explosion diagnostic for self-consistent three-dimensional CCSN simulations.  One may even be able to predict whether a certain
progenitor model will explode without even performing core-collapse
simulations.  This hope will only be realized with further model developments, replacing parameters currently measured from simulations with parameters calculated by other means. 

Before we can even pursue such bold endeavors, however, we
must adapt the integral condition to include the appropriate physics.
For one, we derived the integral condition using Newtonian gravity;
general relativistic considerations are important, therefore we will need to
take the straightforward steps in deriving the condition in GR.
Second, we need to incorporate a more self-consistent neutrino heating
and cooling.  Third, we need to incorporate multidimensional
effects.  Turbulence seems to reduce the critical neutrino
luminosity for explosion; we suspect that one can easily use a
turbulence model to derive a new integral condition for explosion
including turbulence.

In summary, we derived an integral condition for explosion and verified it with one-dimensional parameterized simulations.  When combined with simple steady-state models, it suggests a new explosion diagnostic, $\psimin$, that we argue may be a more predictive measure of a models explodability than other diagnostics.  Finally, we point out that our integral formulation can be extended with better physics, and we are hopeful that this approach may prove useful in disentangling the complicated interactions of various physical effects and in understanding the mechanism of CCSNe in Nature.

\section*{Acknowledgments}
We would like to acknowledge the pioneering work of Adam Burrows,
Hans-Thomas Janka, and Matthias Liebend\"{o}rfer whose contributions to
CCSN theory were directly inspirational to the work presented here.
This material is based upon work supported by the National Science Foundation under Grant No. 1313036.

\newpage
\appendix

\section{Steady-State Equations}
\label{app:steady}

Identifying the steady-state solution with the minimum $\Psi$
is key in developing the explosion diagnostic in section~\ref{sec:explodability} and
Figure~\ref{fig:nearness}.  In this appendix, we therefore present
the steady-state equations that we used to obtain the family of solutions.  

To accommodate the analytic EOS described in the next appendix, we recast the
steady-state equations in terms of the natural independent variables.
Because the analytic EOS in appendix~\ref{app:eos} is most naturally
written as a function of density and temperature, $P = P(\rho, T)$, we rewrite the
steady-state equations as differential equations for $v$,
$\rho$, and $T$.
\begin{equation}
\frac{d \ln v}{d \ln r} = - \left ( 2 + \frac{d \ln
  \rho}{d \ln r}\right ) \,  
\end{equation}
\begin{equation}
\frac{d \ln \rho}{d \ln r} \left ( y P_{\rho} -
\frac{v^2}{\phi}\right ) = 
- y P_T \frac{d \ln T}{d \ln r} - 1 + \frac{2 v^2}{\phi} \, 
\end{equation}
and
\begin{equation}
e_T \frac{d \ln T}{d \ln r} = (\gamma_4 - 1 - e_{\rho})
\frac{d \ln \rho}{d \ln r} - \frac{L_{\nu} \kappa \rho
  r}{\mdot \varepsilon} + \frac{\rho C_0(T/T_0)^6}{\mdot \varepsilon}
\, ,
\end{equation}
where $P_{\rho} = (\partial \ln P / \partial \ln \rho)_{T}$ is
the partial derivative of $P$ with respect to $\rho$ at constant $T$.
Equivalently,
$P_{T} = (\partial \ln P / \partial \ln T)_{\rho}$, 
$e_{\rho} = (\partial \ln \varepsilon / \partial \ln \rho)_{T}$,
and
$e_{T} = (\partial \ln \varepsilon / \partial \ln T)_{\rho}$.

\section{Equation of State}
\label{app:eos}
For the one-dimensional parameterized simulations, we use the tabulated EOS
provided by \citet{hempel12}.  The microphysics includes a distribution of nuclei
that satisfy nuclear statistical equilibrium; the individual
components are, broadly, a dense nuclear component,
ideal gas for the nucleons and isotopes, photon gas, and relativistic electrons and
positrons with arbitrary degeneracy.  The tabulated EOS
and driver are available at \url{http://www.stellarcollapse.org/equationofstate}.

For the steady-state solutions we use an analytic EOS that does
remarkably well in reproducing the microphysics in the region between
the neutrino sphere and the shock.  \citet{bethe90}, \citet{janka01}, and \citet{fernandez09} were useful guides in
developing this analytic EOS.  First, we assume nuclear statistical equilibrium for three
species only: neutrons, protons, and $\alpha$s; their respective mass
fractions are $X_n$, $X_p$, and $X_{\alpha}$.  Conservation of
baryonic mass implies
\begin{equation}
X_n + X_p + X_{\alpha} = 1 \, ,
\end{equation}
and conservation of charge implies
\begin{equation}
X_p + \frac{1}{2} X_{\alpha} = Y_e \,
\end{equation}
where $Y_e$ is the number of electrons per baryon.  In our
steady-state solutions we simply set $Y_e = 0.5$.  In NSE, the Saha
equation provides the remaining equation to find a solution for the abundance of these
three species.
\begin{equation}
X_n^2 X_p^2 = \frac{1}{2}X_{\alpha} \left ( \frac{m_p n_q}{\rho}\right
)^3 \exp(\frac{-Q_{\alpha}}{k_BT}) \, ,
\end{equation}
with
\begin{equation}
n_q = \left ( \frac{m_p k_B T}{2 \pi \hbar^2} \right )^{3/2} \, ,
\end{equation}
and $Q_{\alpha} = 28$ MeV is the binding energy of the $\alpha$.

We construct the pressure and internal energy under the assumptions
that the photons, positrons, and electrons are relativistic and the
partial pressure due to the nucleons and $\alpha$s is given by the
ideal gas law.  In addition, we consider the electrons and positrons with
an arbitrary degeneracy $\eta = \mu_e/(k_B T)$.  The expression for
the degeneracy parameter is
\begin{equation}
\rho = \frac{m_u}{3 \pi^2 Y_e} \left ( \frac{k_B T}{\hbar c } \right
)^3 \eta (\pi^2 + \eta^2) \, .
\end{equation}
The pressure and internal energy may both be divided into partial pressures due to
relativistic ($R$) particles and nonrelativistic ($NR$) particles
\begin{equation}
P = P_R + P_{NR} \, ,
\end{equation}
and
\begin{equation}
\varepsilon = \varepsilon_{R} + \varepsilon_{NR} \, .
\end{equation}
The partial pressure due to the relativistic constituents is
\begin{equation}
P_{R} = \frac{1}{12}\frac{(k_B T)^4}{(\hbar c)^3} \left ( \frac{11
  \pi^2}{15} + 2 \eta^2 + \frac{\eta^4}{\pi^2} \right ) \, ,
\end{equation}
and the partial pressure due to the nonrelativistic constituents is
\begin{equation}
P_{NR} = \left ( 1- \frac{3}{4}X_{\alpha} \right ) \frac{\rho k_B
  T}{m_u} \, .
\end{equation}
The resulting internal energy is
\begin{equation}
\varepsilon = 3\frac{P}{\rho} - \frac{3}{2}\frac{P_{NR}}{\rho} + (1 -
X_{\alpha})\frac{Q_{\alpha}}{4} \, ,
\end{equation}
where the last term is there to account for the transfer of binding
energy per nucleon from $X_{\alpha}$ to the gas.


\end{document}